\documentstyle[11pt,aaspp4,nick]{article}
\begin{document}

\pagestyle{myheadings}
\markright{Gnedin: Cosmological Reionization\hfill}

\title{Cosmological Reionization by Stellar Sources}
\author{Nickolay Y.\ Gnedin}
\affil{CASA, University of Colorado, Boulder, CO 80309;
e-mail: \it gnedin@casa.colorado.edu}


\load{\scriptsize}{\sc}

\def\A{{\cal A}}
\def\B{{\cal B}}
\def\ion#1#2{\rm #1\,\sc #2}
\def\HI{{\ion{H}{i}}}
\def\HII{{\ion{H}{ii}}}
\def\GI{{\ion{He}{i}}}
\def\GII{{\ion{He}{ii}}}
\def\GIII{{\ion{He}{iii}}}
\def\MH{{{\rm H}_2}}
\def\Hp{{{\rm H}_2^+}}
\def\Hm{{{\rm H}^-}}

\def\dim#1{\mbox{\,#1}}

\def\figdir{.}
\def\placefig#1{#1}

\begin{abstract}
I use cosmological simulations 
that incorporate a physically motivated approximation to three-dimensional 
radiative
transfer that recovers correct asymptotic ionization front propagation
speeds for some cosmologically relevant density distributions
to investigate the process of the reionization of the
universe by ionizing radiation from proto-galaxies. Reionization
proceeds in three stages and occupies a large redshift range from
$z\sim15$ until $z\sim5$. During the first, ``pre-overlap'' stage,
$\HII$ regions gradually expand into the low density IGM, leaving behind
neutral high density
protrusions. During the second, ``overlap'' stage, that occurs in about
10\% of the Hubble time, $\HII$ regions merge and the ionizing background
rises by a large factor. During the third, ``post-overlap'' stage, 
remaining high density regions are being gradually ionized as the
required ionizing photons are being produced.

Residual fluctuations in the ionizing background reach significant 
(more than 10\%) levels for the Lyman-alpha forest absorption systems
with column densities above $10^{14}-10^{15}\dim{cm}^{-2}$ at $z=3$ to $4$.
\end{abstract}


\section{Introduction}

Existing ground based observations of the CMB on sub-degree angular
scales suggest that the gas content of the universe was mostly neutral
since recombination at $z\sim1000$ until about
$z\sim100$ (Bond \& Jaffe 1998; Griffiths, Barbosa, \& Liddle 1998), 
because earlier reionization would have
brought the last scattering surface to lower redshift,
smoothing the intrinsic CMB anisotropy. At the
same time we know that the universe is highly ionized since $z\approx5$,
from observations of the spectra of quasars with the highest redshifts
(Giallongo et al.\ 1994; Williger et al.\ 1994; Songaila et al.\ 1999).
This change of the ionization state of the universe from neutral
to highly ionized is called {\it reionization\/}. 

Recent years witnessed a surge in research on reionization along two
separate directions. Semi-analytical methods attempt to describe
the general features of the evolution of the intergalactic medium (IGM)
based on simple, ad hoc assumptions about the star formation history and the
density distribution in the IGM. Most of the previous works utilized a
simple clumping factor to account for the inhomogeneity of the IGM
(Giroux \& Shapiro 1996; Tegmark et al.\ 1997; Madau, Meiksin, \& Rees 1997;
Ciardi \& Ferrara 1997; Haiman \& Loeb 1997, 1998; Madau, Haardt, \& Rees 1999;
Valageas \& Silk 1999; Chiu \& Ostriker 1999). 
A recent work
by Miralda-Escud\'e, Haehnelt, \& Rees (1999) attempted to advance further
by introducing the distribution of the ionized fraction as
a function of gas density which, besides everything else, provides a
specific model for various clumping factors.
Despite using a very simple (perhaps even
oversimplified) ansatz for the ionized fraction - density distribution,
their main conclusions agree remarkably well with the results of
the simulations presented in this paper, as I will show below.

The main advantage of the semi-analytical approaches
is their simplicity and the ability to emphasize the key physical processes.
Their main problem is a significant oversimplification and inability 
to follow the complex dynamical interactions between the dark matter, 
gas, stars, and the spatially inhomogeneous and time-variable
radiation field. Semi-analytical models also miss the effects of correlation
between the sources of ionizing radiation. 

The latter deficiency was partly overcome recently by Ciardi et al.\ (1999)
by combining a semi-analytical model with the $N$-body simulation, but the
lack of dynamics will elude semi-analytical models forever.

Cosmological numerical simulations offer a totally 
different approach to modeling reionization (Ostriker \& Gnedin 1996;
Gnedin \& Ostriker 1997). The main advantage of simulations over the
semi-analytical models is that numerical simulations fully account for
the dynamical evolution of the matter contents of the universe, thus
avoiding the main over-simplification of the semi-analytical models.
Their inherited limitation is a limited dynamic range, which for existing
numerical codes implies an impossibility to achieve numerical convergence
and therefore quantitatively accurate results. However, semi-analytical
models are also imprecise since they adopt ad hoc assumptions and
ansatzes, and thus we must accept the fact that we are not able to
model reionization quantitatively accurate at this moment, and need to
limit ourselves with understanding the key qualitative features of
the physical processes that take place in the universe during that
epochs.

Until recently, numerical simulations failed to include the 
three-dimensional radiative transfer in all its complexity, and therefore
were 
not sufficient to properly model the dynamics of the ionizing radiation
in the universe.

The understanding of the crucial role that the full 3D radiative transfer
plays in the physics of reionization prompted the first attempts to
develop numerical techniques to incorporate the radiative transfer
into numerical simulations (Abel, Norman, \& Madau  1999). 
While Abel et al.\ (1999) were able to develop an algorithm for
an exact implementation of the full 3D radiative transfer into
cosmological simulations, their algorithm is appropriate for following
the radiation field in the case when there are only a few sources
within the volume with the size of the mean free path of an ionizing
photon. When the number of sources within this volume increases,
as is the case after all $\HII$ regions overlap, their method
becomes prohibitively expensive.

In this paper I adopt a somewhat different approach. Rather than trying
to model the radiative transfer exactly, I develop an {\it approximation\/},
which is sufficiently fast computationally. I validate the adopted
approximation on a series of spherically symmetric test cases, for which
the full radiative transfer can be solved exactly at a modest computational
expense.

The physical basis for the adopted approximation, which I call
a ``Local Optical Depth'' (LOD) approximation, is simple. In the
optically thin regime it is straightforward to compute the spatially
variable radiation field, as the problem reduces to simply collecting
$r^{-2}$ potential laws from all the sources in the simulation volume,
which can be done fast with the P$^3$M algorithm. Thus, the key difficulty
in following the full radiative transfer is to account for the optical
depth, which can always be presented as a product of the cross-section,
the gas density, and a characteristic scale. The LOD approximation
adopts an ansatz for this characteristic scale as the scale over
which the gas density changes significantly. 

Having developed the means to follow approximately the 3D spatially 
inhomogeneous
and time-dependent radiative transfer, I apply them to modeling the process
of reionization of the universe in a specific scenario when sources of
ionizing radiation are stars (grouped into proto-galaxies). There are two
reasons why numerical simulations need to be restricted to this
specific case at the moment.

The first reason is ``scientific'', as the mounting evidence suggests
that quasars are unable to reionize the universe at $z>5$
(Madau et al.\ 1999), and thus the universe is reionized by stars.
The second reason is ``pragmatic''. A quasar radiation field may affect
the properties of the IGM over a region several tens of megaparsecs in size.
Thus, a numerical simulation capable of modeling the reionization by quasars
would have to have a computational region of at least $100\dim{Mpc}$ in size.
At the same time, in order to resolve the formation of the first structures
and to follow the gas density with sufficient precision,
it will need spatial resolution below $1\dim{kpc}$ and mass resolution
better than about $10^6$ solar mass in baryons. Such a dynamic range
($10^5$)
is beyond the capabilities of the existing cosmological hydrodynamic 
codes.\footnote{With the exception of the Adaptive Mesh Refinement
technique.}

Thus, only a numerical simulation with the box size of several
Mpc is practical at the moment. Fortunately, this box size is 
sufficient to model the reionization by stellar sources (on a semi-qualitative 
level), as will be
shown below.

This paper is organized in a conventional way. In \S 2 I describe the
simulations, \S 3 is devoted to the results, and \S 4 concludes with
the discussion. Appendix describes the Local Optical Depth approximation
and the tests of the method.

\section{Simulations}

Simulations reported in this paper were performed with the 
``Softened Lagrangian Hydrodynamics'' (SLH-P$^3$M) code (Gnedin 1995, 1996;
Gnedin \& Bertschinger 1996). The following physical ingredients are
currently included in the code:
\begin{description}
\item[Dark matter] is followed using the adaptive P$^3$M algorithm.
\item[Gas dynamics] is followed on a quasi-Lagrangian deformable mesh
using the SLH algorithm.
\item[Star formation] is included using the Schmidt law in resolution
elements that sink below the numerical resolution of the code.
\item[Atomic physics] of hydrogen and helium plasma is followed exactly
using a two-level implicit scheme.
\item[Molecular hydrogen] formation and destruction is followed
exactly (including the radiative transfer effects)
in the limit when the fraction of hydrogen in the molecular
form is small and the self-shielding of $\MH$ is unimportant
(in the latter case the approximate method of following the
radiative transfer becomes exact).
This is always the case in the simulation presented in this
paper because the numerical resolution is not sufficient to resolve
the formation of molecular clouds.
\item[Radiative transfer] is treated self-consistently, albeit only
approximately, in a 3D
spatially-inhomogeneous and time-dependent manner. This is a new
addition to the SLH-P$^3$M code. The specific implementation of the
radiative transfer is described in the Appendix.
\end{description}

Hereafter I adopt the following values of cosmological parameters
in the CDM+$\Lambda$ model:
$$
        \Omega_0 = 0.3,\ \ \ \Omega_\Lambda = 0.7,\ \ \ h = 0.7,\ \ \ 
        \Omega_b = 0.04.
$$
This cosmological model is in a reasonable agreement with all available
observational data and has a desirable feature that all the parameters
are specified to only one decimal place. For the simulations presented
in this paper the particular choice of cosmological parameters matters
rather little as all reasonable models look similar at high redshift.

There are two free parameters that control the behavior of the simulations.
The first parameter is the efficiency of star formation $\epsilon_*$, that
appears in the Schmidt law:
\begin{equation}
	{d\rho_*\over dt} = \epsilon_* {\rho_g\over t_*},
	\label{sfeq}
\end{equation}
where $\rho_*$ and $\rho_g$ are the stellar and the gas density
respectively, and 
$t_*$ is the maximum of the dynamical and cooling time. Equation (\ref{sfeq})
is only applied in resolution elements that are determined to be beyond
the resolution limit of the simulation. In all resolved resolution elements
star formation is not allowed.

The second important parameter measures the amount of ionizing radiation
each ``star'' emits:
\begin{equation}
	{du_{UV}\over dt} = \epsilon_{UV} c^2 {d\rho_*\over dt},
	\label{uuveq}
\end{equation}
where $u_{UV}$ is the production of the
energy density in ionizing radiation per unit time. This parameter apparently
depends on the initial mass function (IMF). For the Salpeter IMF the
``escape fraction'' of the ionizing photons, i.e.\ the fraction of
ionizing photons that escape from the immediate vicinity of a star, is
roughly
\begin{equation}
	f_{\rm esc} \approx 1.4\times10^4\epsilon_{UV}
	\label{escfrac}
\end{equation}
(Madau et al.\ 1999). Thus, an efficiency of $4\times10^{-5}$ corresponds
to roughly 60\% escape fraction.

However, what matters most for the evolution of the IGM is only the total
number of ionizing photons emitted, i.e.\ the product of $\epsilon_*$
and $\epsilon_{UV}$. Therefore, 
I fix $\epsilon_*=0.05$ as my fiducial value, 
which is similar to the values found in the Milky Way.
The latter by itself is not a justification for this particular choice,
since there may be very little in common between the star formation in the
Milky Way and at high redshift. 
However, this choice leads to about
3 to 5 percent of all baryons being converted to stars by $z=4$, which
translates to about 10 to 20 percent of all stars being formed by
$z=4$, a reasonable estimate given the current data on the star formation
history of the universe (Madau 1999, Steidel et al.\ 1999, Renzini 1999).
In addition, the star formation rate at $z=4$ in the simulation turns
out to be similar to the observational value (see Fig.\ \ref{figSF} below).

It is important to emphasize here that an assumed value of the radiation
efficiency, or for that matter of the escape fraction,
is resolution dependent, since it measures the fraction of
energy that escapes from simulated stars into the simulated gas. Since
any simulation has a finite resolution, this means that the escape fraction
in a simulation measures the fraction of radiation that ``escapes from the
stellar surface to the resolution scale of the simulation'', and 
mathematically precise meaning of this phrase cannot be formulated, as
it may depend on the behavior of a numerical scheme at its resolution
limit. It may even be a bad approximation to assume $\epsilon_{UV}$ to be
constant during the simulation, since simulations presented in this paper
have a fixed resolution in comoving coordinates, which means that in physical
units the resolution length increases with time. However, since the process of
reionization takes a relatively short period of time, this effect is
likely to be insignificant. At any rate, as I argue below, the simulations
presented in this paper are only reliable on a semi-qualitative level (within a
factor of two or so), and
cannot be relied upon to provide an accurate quantitative answer.

\def\tableone{
\begin{deluxetable}{cccccccc}
\tablecaption{Numerical Parameters\label{tabone}}
\tablehead{
\colhead{Run} & 
\colhead{$N$} & 
\colhead{Box size} & 
\colhead{Baryonic mass res.} & 
\colhead{Spatial res.} & 
\colhead{$\epsilon_*$} & 
\colhead{$\epsilon_{UV}$} & 
\colhead{$z_f$} }
\startdata
N64\_L2\_A & $64^3$ & $2h^{-1}{\rm\,Mpc}$ & $10^{5.7}\dim{M}_{\sun}$ & 
$1.5h^{-1}{\rm\,kpc}$ & 0.05 & $4\times10^{-5}$ & 4 \\
N64\_L2\_B & $64^3$ & $2h^{-1}{\rm\,Mpc}$ & $10^{5.7}\dim{M}_{\sun}$ & 
$1.5h^{-1}{\rm\,kpc}$ & 0.05 & $1.2\times10^{-4}$ & 4 \\
N64\_L2\_C & $64^3$ & $2h^{-1}{\rm\,Mpc}$ & $10^{5.7}\dim{M}_{\sun}$ & 
$1.5h^{-1}{\rm\,kpc}$ & 0.05 & $1.3\times10^{-5}$ & 4 \\
N64\_L2\_D & $64^3$ & $2h^{-1}{\rm\,Mpc}$ & $10^{5.7}\dim{M}_{\sun}$ & 
$1.5h^{-1}{\rm\,kpc}$ & 0.10 & $4\times10^{-5}$ & 4 \\
 & & & & & & \\
N128\_L4\_A & $128^3$ & $4h^{-1}{\rm\,Mpc}$ & $10^{5.7}\dim{M}_{\sun}$ & 
$1.0h^{-1}{\rm\,kpc}$ & 0.05  & $4\times10^{-5}$ & 4 \\
N128\_L8\_A & $128^3$ & $8h^{-1}{\rm\,Mpc}$ & $10^{6.6}\dim{M}_{\sun}$ & 
$2.0h^{-1}{\rm\,kpc}$ & 0.05  & $4\times10^{-5}$ & 6.5 \\
N128\_L2\_A & $128^3$ & $2h^{-1}{\rm\,Mpc}$ & $10^{4.8}\dim{M}_{\sun}$ & 
$0.5h^{-1}{\rm\,kpc}$ & 0.025 & $4\times10^{-5}$ & 6.5 \\
\enddata
\end{deluxetable}
}
\placefig{\tableone}
Table \ref{tabone} lists all the simulations presented in this paper. The last
column in the table is the final redshift at which a given simulation is 
stopped. The
first four simulations whose labels begin with N64 contain $64^3$ dark matter
particles, $64^3$ baryonic mesh, and a number of stellar particles which
form during the simulation. These runs have a dynamic range of 1300. 
All $64^3$ runs have a computational box size of $2h^{-1}$ comoving
megaparsecs. 

The $64^3$ runs serve as tests and designed to investigate the parameter
dependence and the convergence of the simulations. They are not used for
producing scientific results. The first three of them differ only
by the value of the radiation efficiency, and the last run, N64\_L2\_D,
has a twice higher star formation rate. One may note that the escape 
fraction in run N64\_L2\_B is formally 170\% if equation (\ref{escfrac})
is used. This should not be considered unphysical since this run is
only used as a test, and besides that a 70\% increase in the fraction
of high mass stars will compensate for that. Anyway, as I show below,
this value of the radiation efficiency does not fit the observations.

The last three simulations listed in Table \ref{tabone} whose labels begin
with N128 incorporate $128^3$ dark matter particles, the same number of
baryonic cells, and about twice more stellar particles, which keep forming
during the simulation. The dynamic range of these three simulations is 4000,
and other parameters are listed in the table. Run N128\_L4\_A is the
production run of this paper, and two other $128^3$ runs are also used to
investigate numerical convergence, and therefore are not continued until
$z=4$.

\def\capRS{
Mass and spatial resolution of the simulations presented in Table 
\protect{\ref{tabone}}. Left $y$ axis labels comoving scales and
right $y$-axis labels physical scales at $z=10$.
The thin solid box shows all four $64^3$ runs
(slightly offset for clarity),
the bold solid box shows run N128\_L4\_A, and two bold dashed boxes
show runs N128\_L2\_A and N128\_L8\_A. 
The shaded regions marks the mass
scales on which baryons do not cluster in linear theory due to finite
pressure (the so-called filtering scale; Gnedin \& Hui 1998).}
\placefig{
\begin{figure}
\epsscale{0.70}
\insertfigure{\figdir/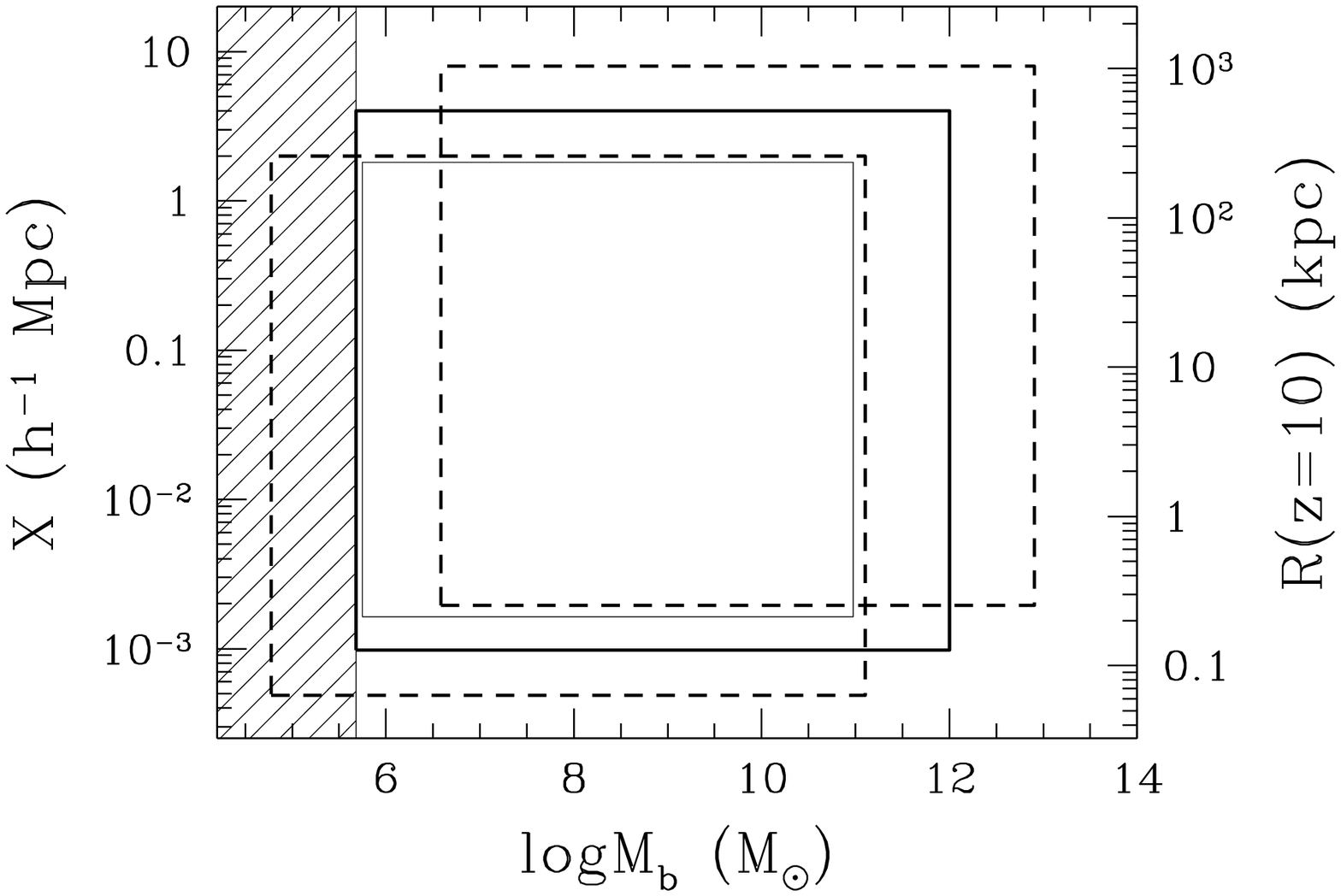}
\caption{\label{figRS}\capRS}
\end{figure}
}
Figure \ref{figRS} illustrates in a graphical form the mass and spatial
scales simulated in this paper. The production run N128\_L4\_A has just
enough mass resolution to resolve the filtering scale, the minimum scale
on which baryons cluster in linear theory.\footnote{As has been shown in 
Gnedin
\& Hui (1998) at $z=10$ the mass corresponding to this scale is about 
11 times larger than the Jeans mass at this redshift.} This is clearly
only marginally enough, as one would prefer to resolve the filtering
scale by a factor of 10 or more to have a safe margin. On the other hand
the production run N128\_L4\_A does not miss a large amount of small
scale power. Thus, even from this consideration we should expect that
the simulations presented in this paper are not quantitatively accurate,
but should be sufficient to give a general qualitative picture, since
they include all the relevant physics and account for most of the 
initial power.

\def\capSF{
The evolution of the spatially averaged ionizing intensity $J_{21}$
(measured in conventional units of 
$10^{-21}\dim{erg}/\dim{cm}^2/\dim{sec}/\dim{Hz}/\dim{rad}$) 
({\it a\/})
and the star formation rate ({\it b\/}) as a function of redshift
for seven runs listed in Table \protect{\ref{tabone}}: 
N64\_L2\_A ({\it thin short-dashed line\/}),
N64\_L2\_B ({\it thin long-dashed line\/}),
N64\_L2\_C ({\it thin dotted line\/}),
N64\_L2\_D ({\it thin solid line\/}),
N128\_L4\_A ({\it bold solid line\/}),
N128\_L8\_A ({\it bold long-dashed line\/}),
N128\_L2\_A ({\it bold dotted line\/}). The data points for the star
formation rate at $z\approx4.2$ are from Steidel et al.\ (1999; 
{\it filled circle\/}) and the same data corrected for dust extinction
by a different method by Nagamine, Cen, \& Ostriker (1999;
{\it empty circle\/}).
}
\placefig{
\begin{figure}
\epsscale{0.65}
\inserttwofigures{\figdir/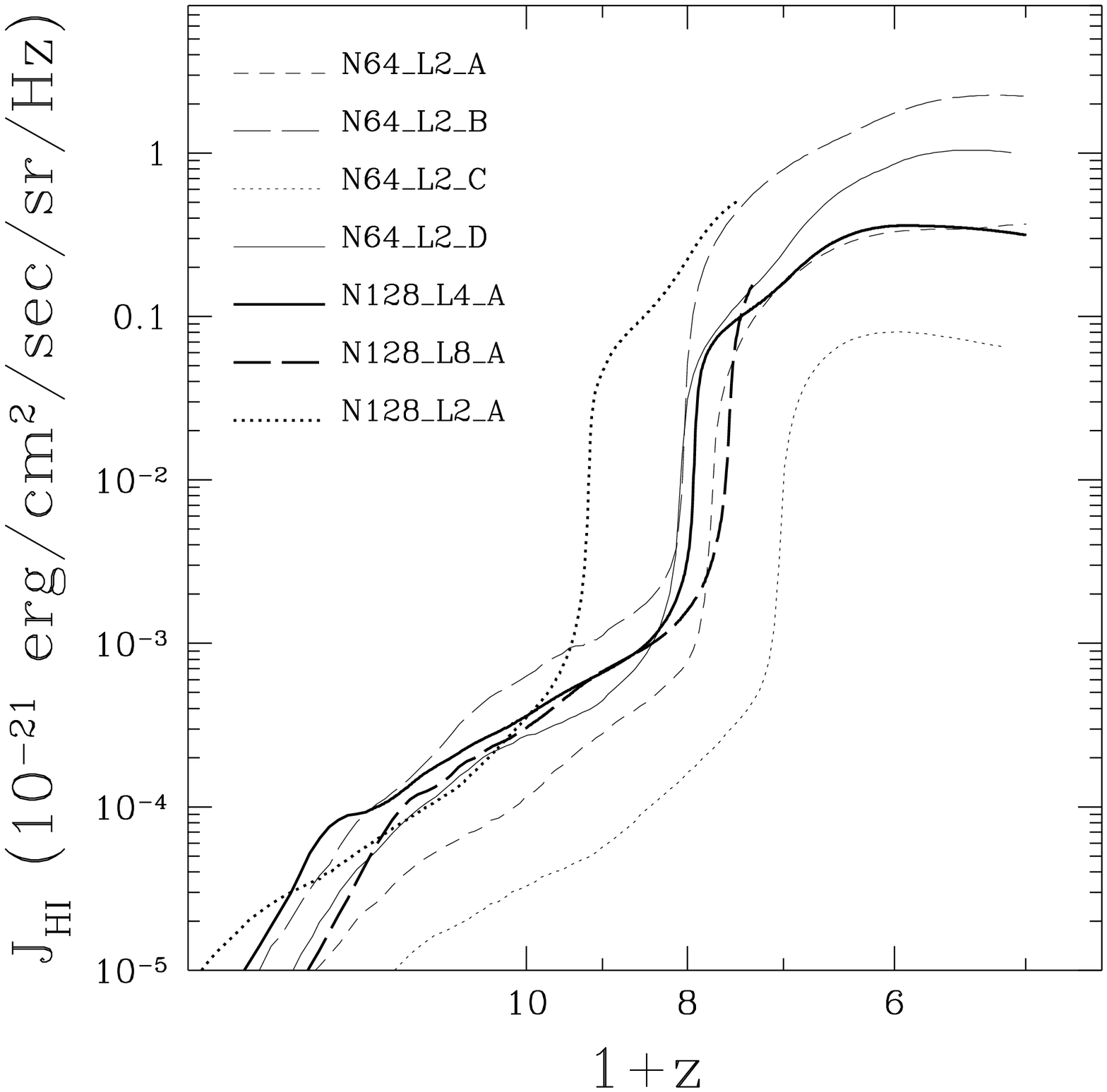}{\figdir/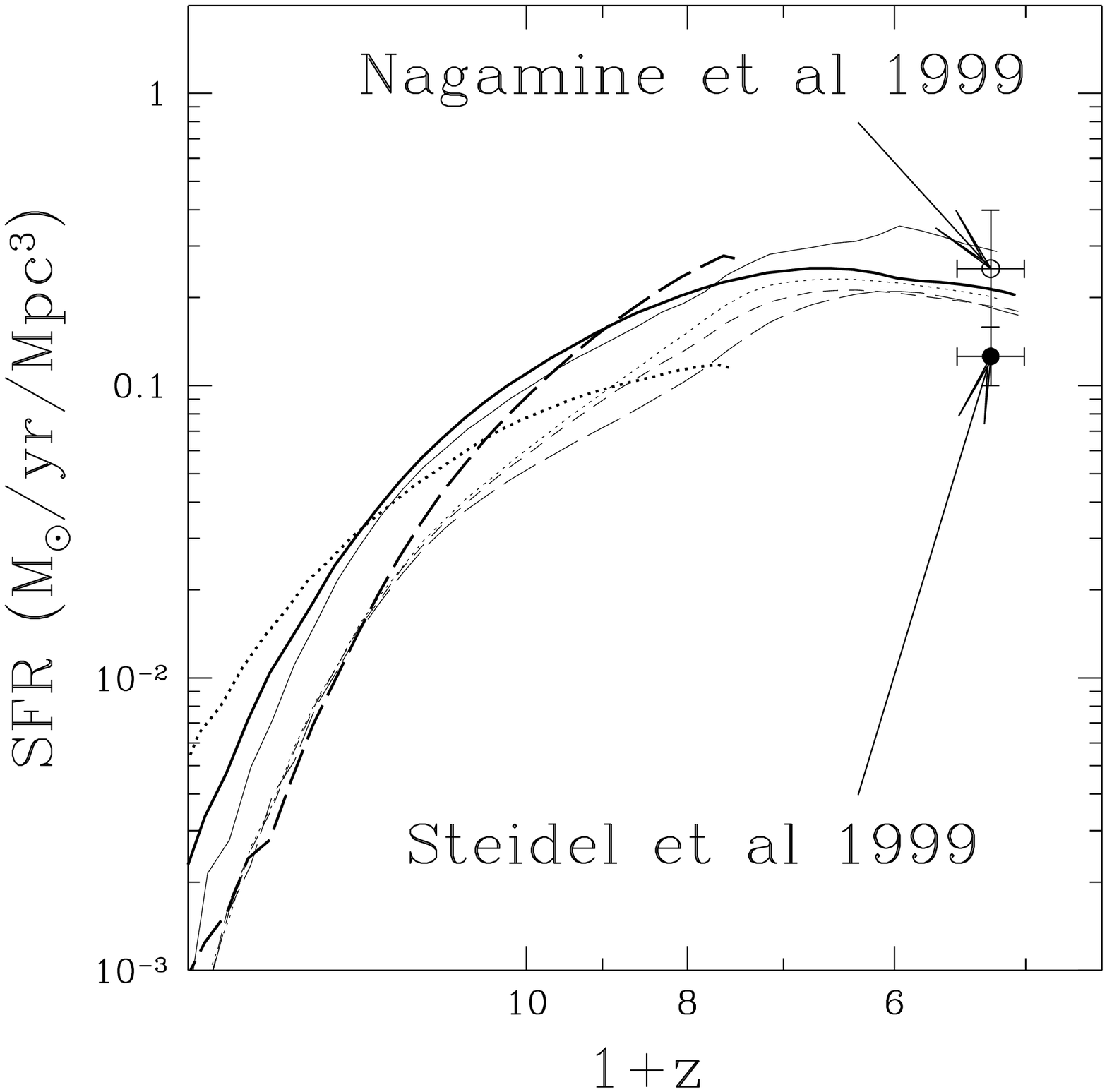}
\caption{\label{figSF}\capSF}
\end{figure}
}
Figure \ref{figSF} now demonstrates the level of convergence of the
simulations presented in this paper and the method of choosing the
parameter $\epsilon_{UV}$. Three thin lines: dotted, short-dashed, and
long-dashed
show runs N64\_L2\_C, N64\_L2\_A, and N64\_L2\_B respectively, 
which only differ by the value of $\epsilon_{UV}$. One can see that
at $z=4$ the value of the ionizing intensity $J_{21}$
is 0.06, 0.4, and 2.5,
respectively, changing by a factor of about 6.5 whenever the
radiation efficiency changes by a factor of 3. Since the observational
values for $J_{21}$ at $z=4$ range between 0.2 and 0.4 (Lu et al.\ 1996),
I choose $\epsilon_{UV}=4\times10^{-5}$ (used in run N64\_L2\_A) as a
reasonable value for this parameter. Since my simulations cannot be used for
accurate quantitative predictions, such a rough agreement with the 
observations is sufficient.

A similar exercise with $128^3$ simulations is however not feasible, as
it would require an excessive amount of computer time to run
three $128^3$ simulations up to $z=4$ with three different values of
$\epsilon_{UV}$. However, as can be seen from the top panel of
Fig.\ \ref{figSF}, the production run simulation N128\_L4\_A with
assumed value of $\epsilon_{UV}=4\times10^{-5}$ also gives $J_{21}=0.4$
at $z=4$, and thus a guess based on $64^3$ simulations turns out to
be a good one, whereas a small run N64\_L2\_D, which has twice higher
star formation rate than N64\_L2\_A, agrees better with the production
run before and during the epoch of overlap
(indicated by a sharp rise
in $J_{21}$).

The two other $128^3$ runs were normalized so as to agree with the 
production run N128\_L4\_A at $z\sim10$ in the star formation rate
and the ionizing intensity $J_{21}$ (the small box run N128\_L2\_A
actually has somewhat lower star formation rate). Nevertheless, they
disagree with it at the epoch of overlap,
and this difference indicates the lack of numerical
convergence. 

Let me first focus on an $8h^{-1}\dim{Mpc}$ run N128\_L8\_A. It agrees
quite well with the production ($4h^{-1}\dim{Mpc}$) run before the 
overlap, but predicts a somewhat lower redshift of overlap. It however
agrees better with the smaller N64\_L2\_A run, because its spatial resolution
is closer to the small run than to the production run. Thus, the redshift
of overlap is delayed if the spatial resolution of a simulation is not
sufficient. Is the resolution of the production run sufficient in this
respect? We can use comparison between the production run N128\_L4\_A
and a $2h^{-1}\dim{Mpc}$ run N128\_L2\_A to investigate the effect of
spatial resolution. It is clear that the production run still lacks
resolution to resolve the very first star formation at $z\sim 20$. This
is not surprising, since it barely resolves the baryon filtering scale
before the reheating, and thus misses about 7\% of the total small scale
power. At $z\sim10$ the two runs more-or-less agree, but the smaller run
has a much earlier time of overlap. Also, it gives $J_{21}$ at $z=6.5$
of about 0.6, already higher than the final value of the production run
of about 0.4. Thus, run N128\_L2\_A can be considered as giving
the upper limit
for the redshift of overlap for this model (which is $z=8.5$), and 
this upper limit is likely to be too high, because
run N128\_L2\_A is going to have a too high a value for $J_{21}$ at
lower redshifts. This is of course not surprising, since run N128\_L2\_A
has the same box size as N64\_L2\_A, but much higher spatial and mass
resolution, so it is going to have more star formation than a smaller
run, and thus an earlier reionization. In comparison to the production
run N128\_L4\_A, it misses important large scale waves which contribute
toward absorption. It also suffers from the cosmic variance
problem, since at $z=4$ the rms density fluctuation in a $2h^{-1}\dim{Mpc}$
box is 0.41, whereas for a $4h^{-1}\dim{Mpc}$ it is 0.27, and for a
$8h^{-1}\dim{Mpc}$ it is 0.16.

The comparison above does not however constitute the complete test of
numerical convergence, as three different resolutions are required for such
a test. Thus, a $256^3$ run is needed to assign a meaningful value
for the numerical error of my simulations, but such a run is currently
beyond the available means. I must therefore acknowledge that the
simulations presented in this paper can only be considered reliable on a 
semi-qualitative basis, within a factor of two or so. One can use the
difference between the bold solid line and thin solid and dashed lines
as an estimate of the numerical error, but without a larger simulation it
is impossible to assign any meaningful confidence level to this error.

\section{Results}

\subsection{Reionization at a glance\label{sec:rg}}

\def\capIM{
A thin slice through the simulation volume at eight different epochs:
({\it a\/}) $z=11.5$, 
({\it b\/}) $z=9$, 
({\it c\/}) $z=7.7$, 
({\it d\/}) $z=7$, 
({\it e\/}) $z=6.7$, 
({\it f\/}) $z=6.1$, 
({\it g\/}) $z=5.7$, 
({\it h\/}) $z=4.9$. Shown are logarithm of 
neutral hydrogen ({\it upper-left\/}),
logarithm of gas density ({\it lower-left\/}),
logarithm of gas temperature ({\it lower-right\/}),
and $\lg(J_{21})$ as a function of redshift ({\it upper-right\/}). 
}
\placefig{
\begin{figletters}
\begin{figure}
\epstwoscale{0.40}
\insertfourfigures{\figdir/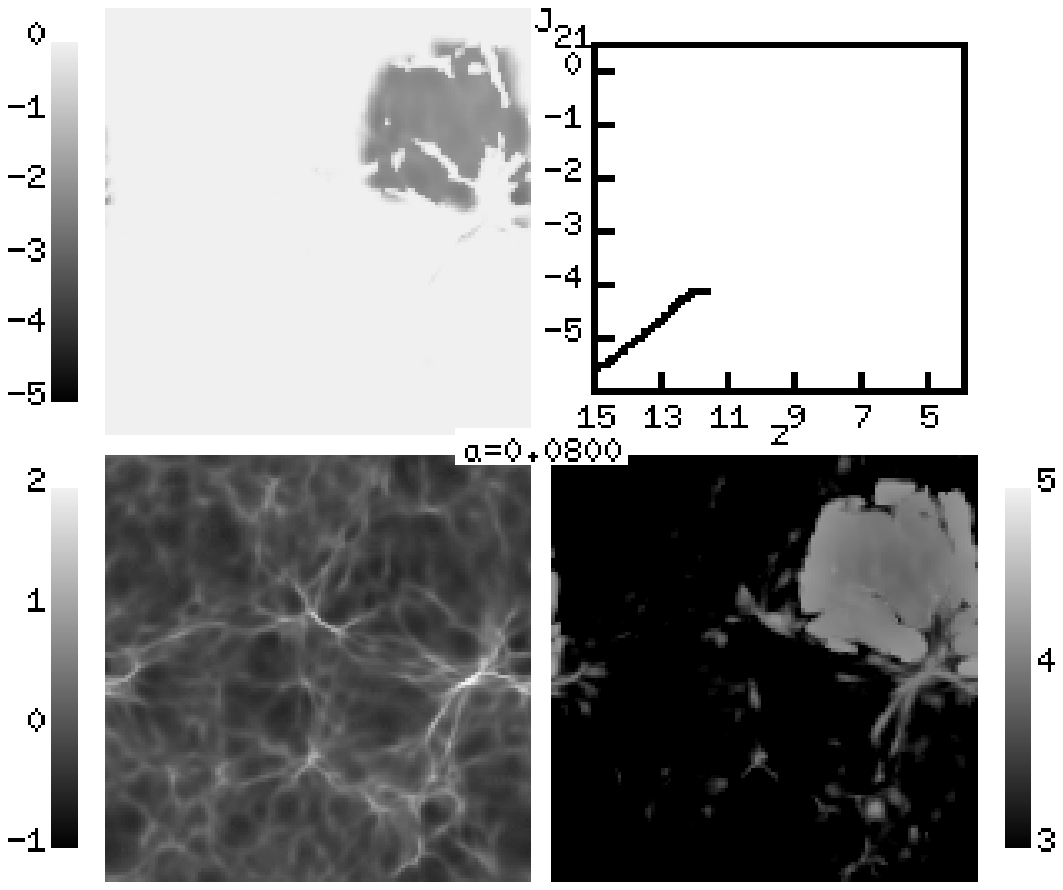}{\figdir/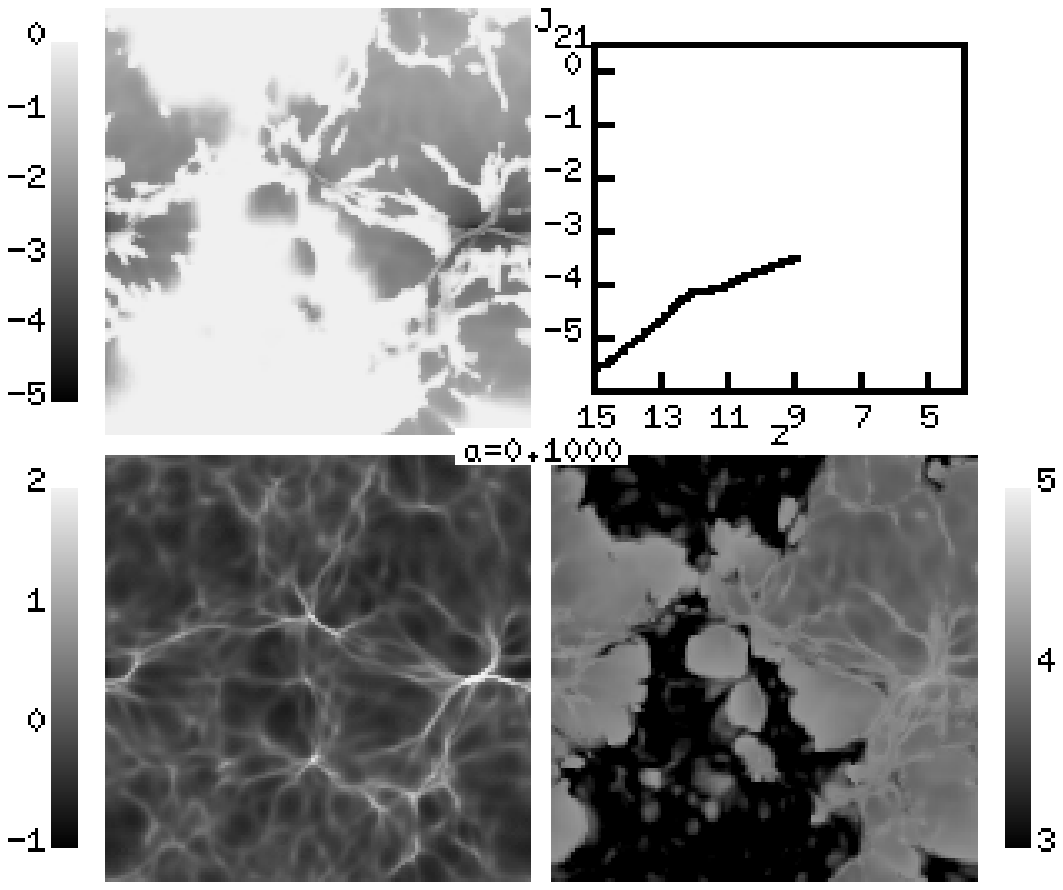}{\figdir/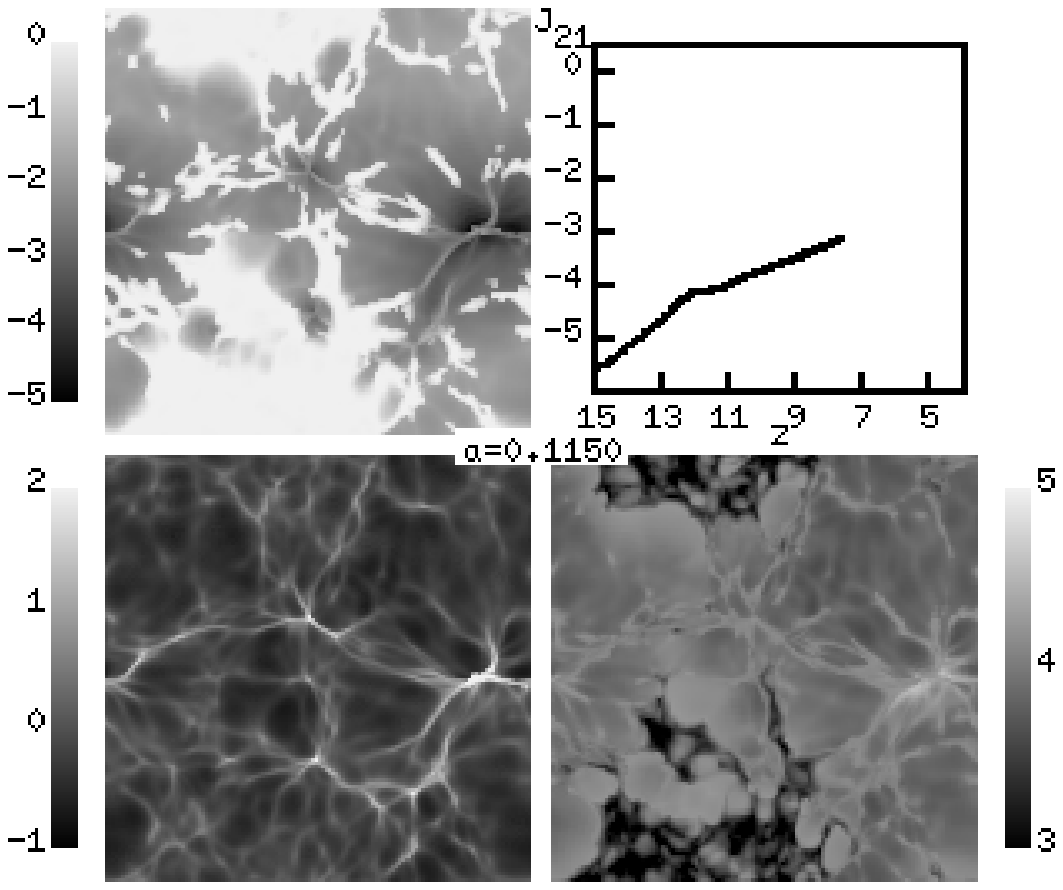}{\figdir/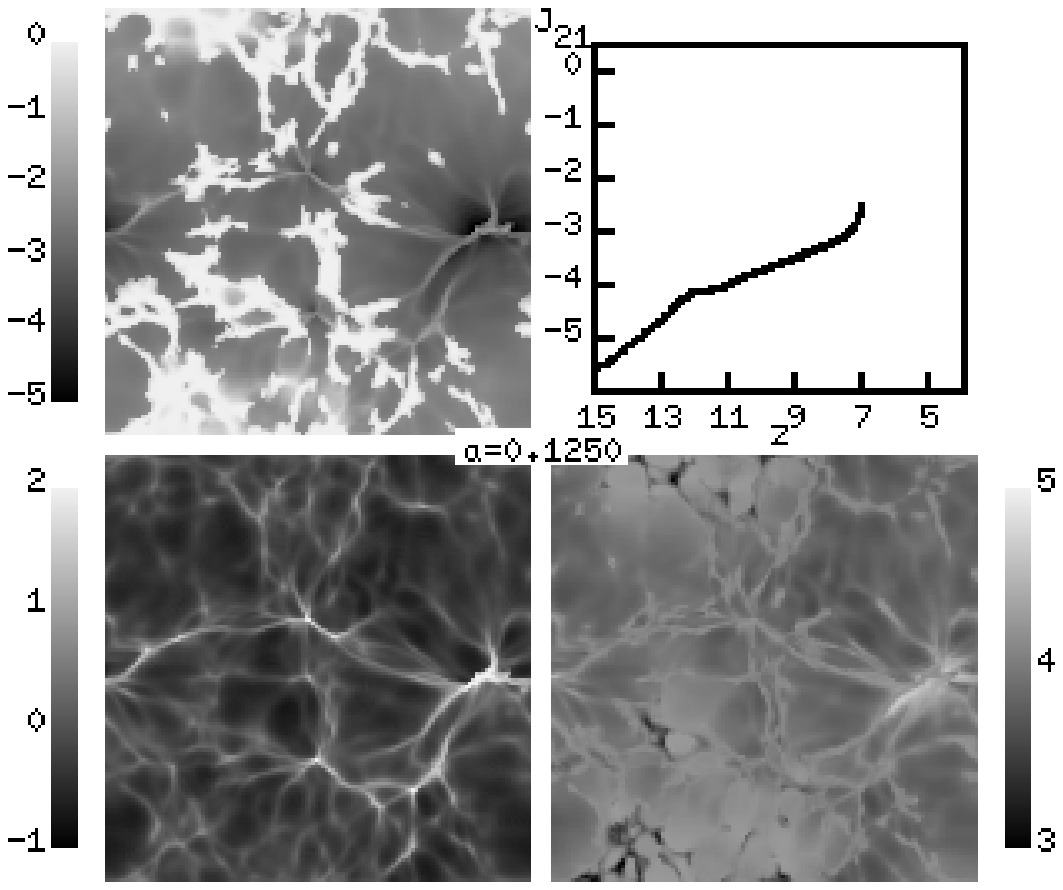}
\caption{\label{figIM}\capIM}
\end{figure}
}
\placefig{
\begin{figure}
\epstwoscale{0.40}
\insertfourfigures{\figdir/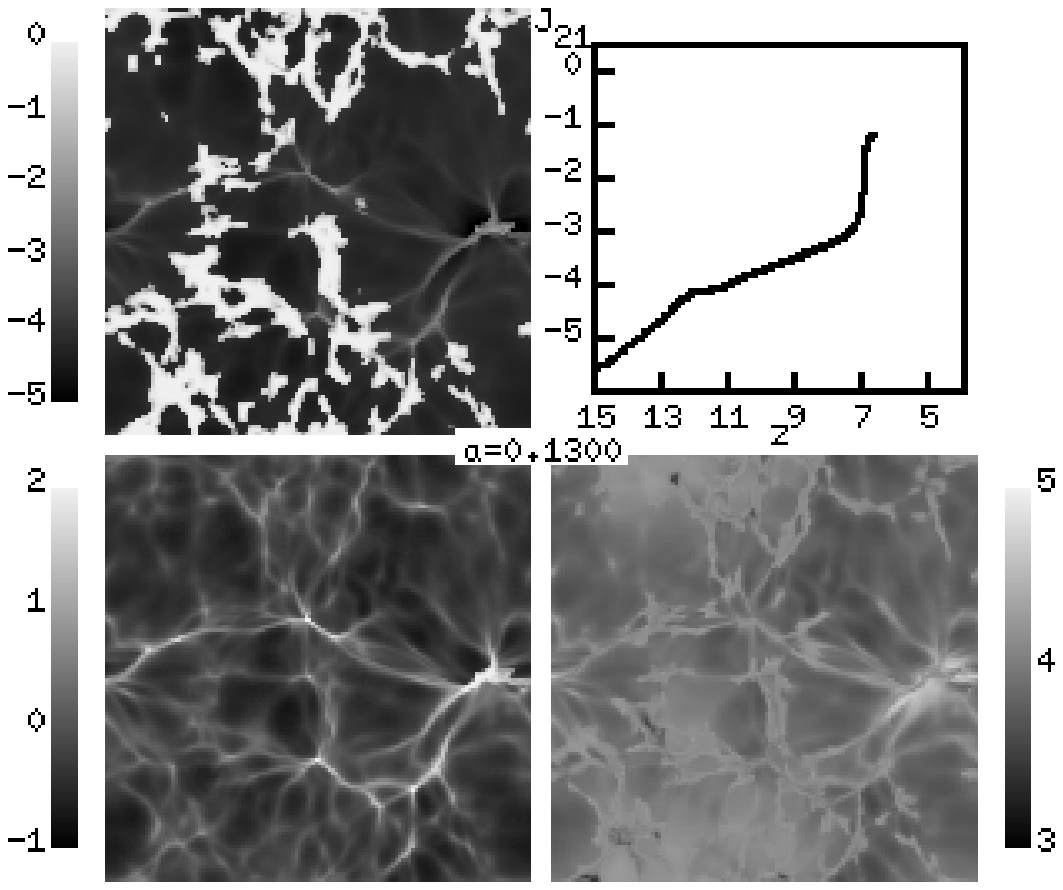}{\figdir/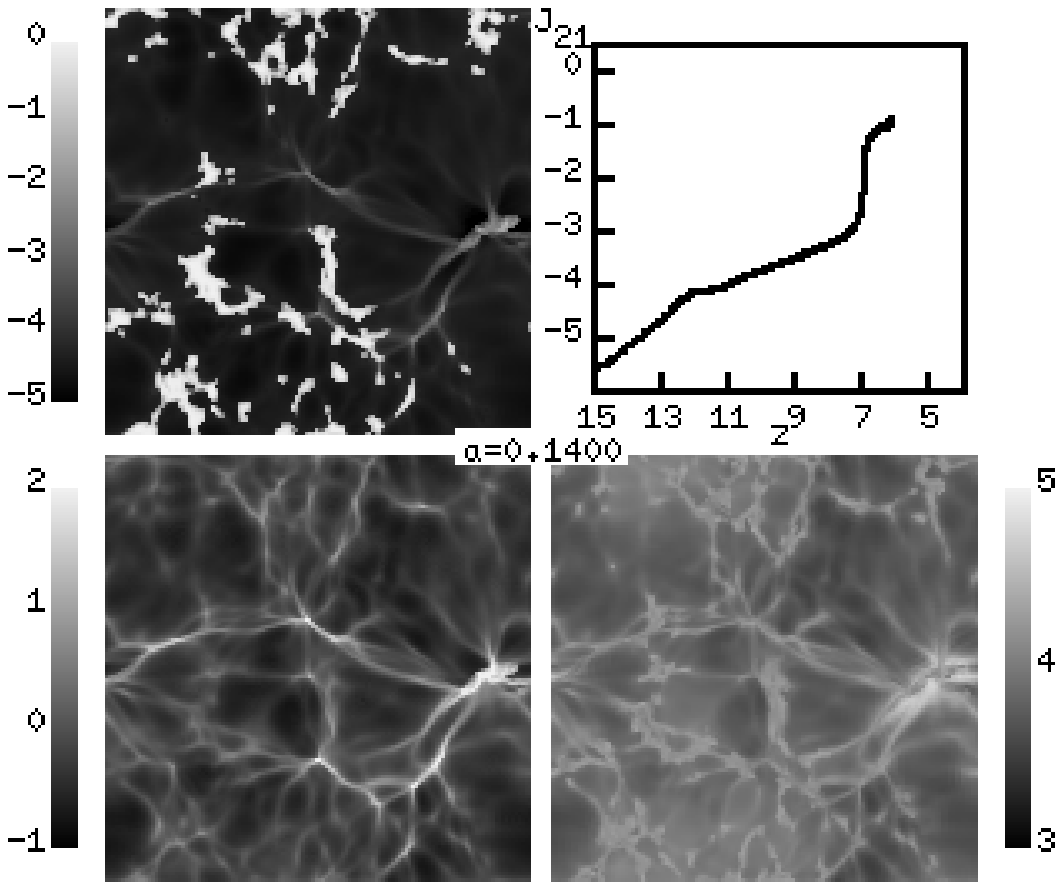}{\figdir/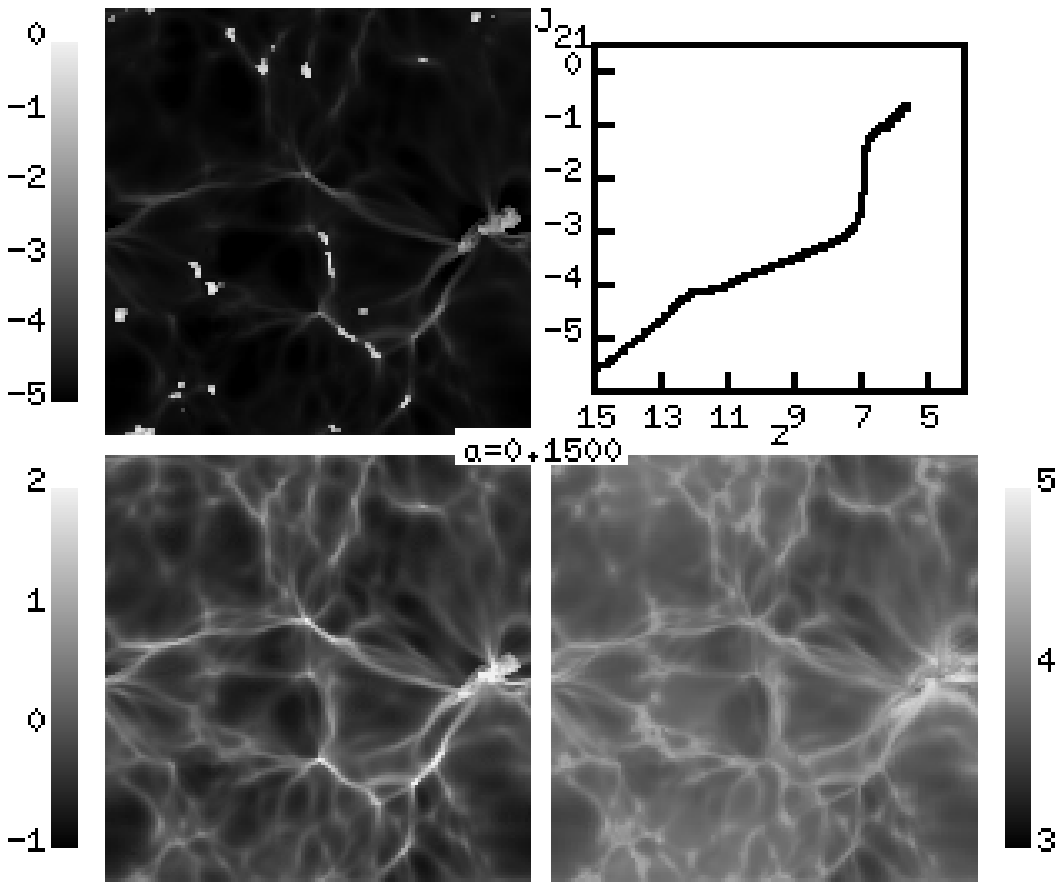}{\figdir/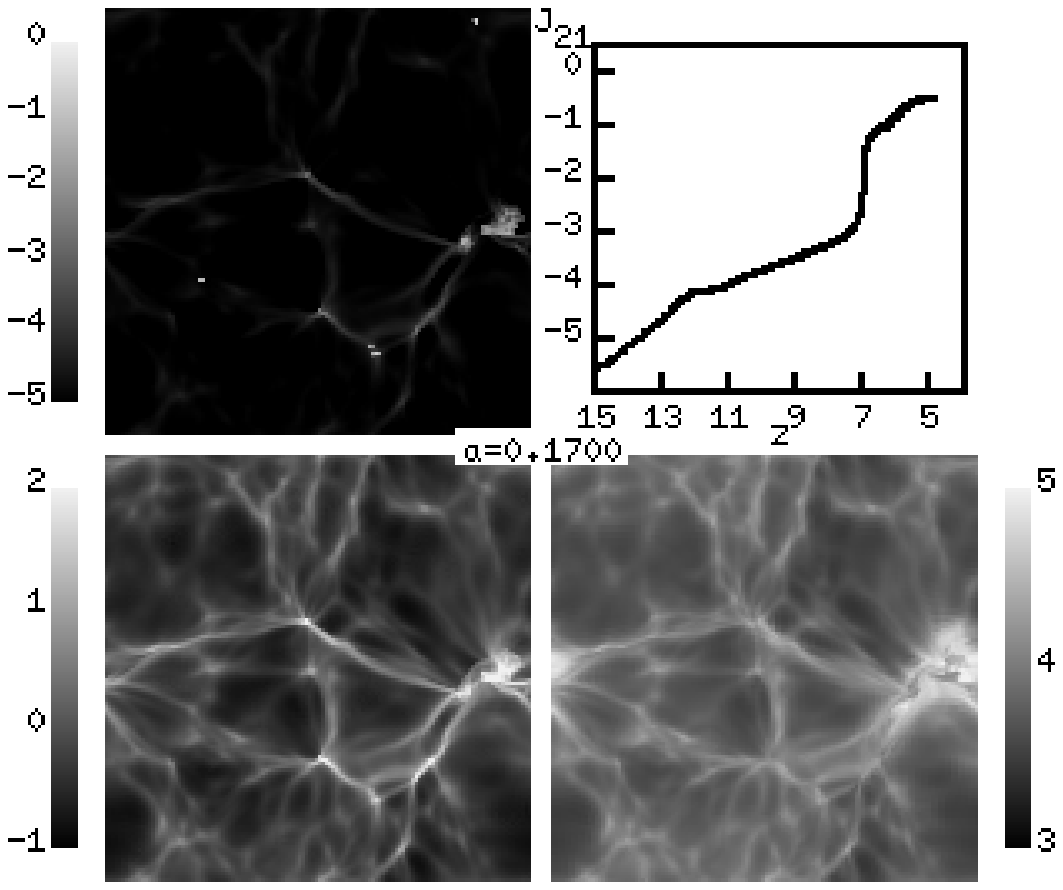}
\end{figure}
\end{figletters}
}
We can now look at a general description of the reionization process,
as reflected in the simulations. In order to visualize the process,
I show in Figure \ref{figIM} a thin ($15 h^{-1}$ comoving kiloparsecs
deep) slice through the computational box taken at a random place.
The three panels show in a logarithmic stretch the neutral hydrogen
(upper left panel), the gas density (lower left panel), and the gas
temperature (lower right panel). The upper right panel shows the evolution
of the ionizing intensity (logarithmically scaled) versus the redshift
(the same as the solid bold line in Fig.\ \ref{figSF}a). The scale bars
next to respective panels show the correspondence between the color and
the numeric scale. The epoch as measured by the scale factor
$a\equiv1/(1+z)$ is labeled at the center of each 
plot.\footnote{An MPEG movie of the simulation is available at
{\tt http://casa.colorado.edu/$\sim$gnedin/GALLERY/rei\_p.html}.}

I can now describe the general features of reionization. The reionization
starts (Fig.\ \ref{figIM}a) with ionization fronts propagating from 
proto-galaxies located in high
density regions into the voids, leaving the high density outskirts of an object
still neutral, because at high density the recombination time is very 
short, and there are not enough photons to ionize the high density regions.
As an ionization front moves on with the $\HII$ region forming behind it, it 
leaves behind high density regions which require many more photons to get
ionized than is available at that moment (Fig.\ \ref{figIM}a-d). This stage
of the reionization process can be called ``pre-overlap'', and it extends
over a considerable range of redshifts $\Delta z\sim 5$ around $z\sim 10$.
During this time the high density regions around the source are slowly 
becoming ionized, whereas high density regions far from the source remain
neutral. The ionizing intensity at this time remains low and is slowly
increasing with time. However, since the radiation field is highly 
inhomogeneous at this time, the ionizing intensity $J_{21}$, which is by
definition a space-average, has only a formal meaning.

By $z\approx 7$ the $\HII$ regions start to overlap (Fig.\ \ref{figIM}d),
with the result that the number of sources shining at an average place in the 
universe increases, and the ionizing intensity starts to rise rapidly.
The process of reionization enters its second stage, the ``overlap'',
which is quite rapid (Fig.\ \ref{figIM}d-e).
As the ionizing intensity is rapidly increasing,
the last remains of the neutral low density IGM are quickly eliminated,
the mean free path increases by some two orders of magnitude (as explained
below) over a Hubble time or so, and voids become highly ionized
(neutral fraction of the order of $10^{-5}$). The high density regions at
this moment are still neutral, as the number of ionizing photons available
is not sufficient to ionize them.

After the overlap is complete, the universe is left with highly ionized 
both low 
density regions and some of the high density ones
(which happened to lie close to
the source, where the local value of the ionizing intensity is higher
than the spatial average). High density regions far from any source
remain neutral. This stage can be called ``post-overlap'', and is well
described by the semi-analytic model of Miralda-Escud\'{e} et al.\ (1999).
As time goes on, and more and more ionized photons are emitted, the high 
density regions are gradually being eaten away (Fig.\ \ref{figIM}e-g), 
and the spatially averaged
ionizing intensity $J_{21}$ continues to rise slowly until it flattens
out at $z\sim 5$ (Fig.\ \ref{figIM}g). 
The latter effect is however likely to be an artifact
of a finite simulation box, as by this time the mean free path exceeds the
box size by about a factor of 10 (as is shown below), 
and the simulation runs out of high
enough density peaks, which would be present in the real universe (or a 
larger box simulation), and fails to reproduce the reality even on a
semi-qualitative level.

Therefore, the recent argument of whether reionization is ``fast'' or
``slow'' (see, for example, Miralda-Escud\'{e} et al.\ 1999 versus
Gnedin \& Ostriker 1997 or Madau et al.\ 1999) is, in large part, a
question of terminology. If one considers the whole process of reionization,
which consists of ``pre-overlap'', ``overlap'', and ``post-overlap'',
one inevitably concludes that reionization is slow (i.e.\ taking place
over a Hubble time or longer). However, if one looks only at the process
of ``overlap'' and labels it reionization, then one concludes that
reionization is fast.

\subsection{Reionization in more detail}

\def\capMP{
The evolution of the mean free path ({\it a\/}) and its rate of change
({\it b\/}) for the production run N128\_L4\_A ({\it bold solid line\/})
and two smaller runs N64\_L2\_D ({\it thin solid line\/}) and
N64\_L2\_A ({\it thin short-dashed line\/}). Two thin dotted lines are
explained in the text.
}
\placefig{
\begin{figure}
\epsscale{0.65}
\inserttwofigures{\figdir/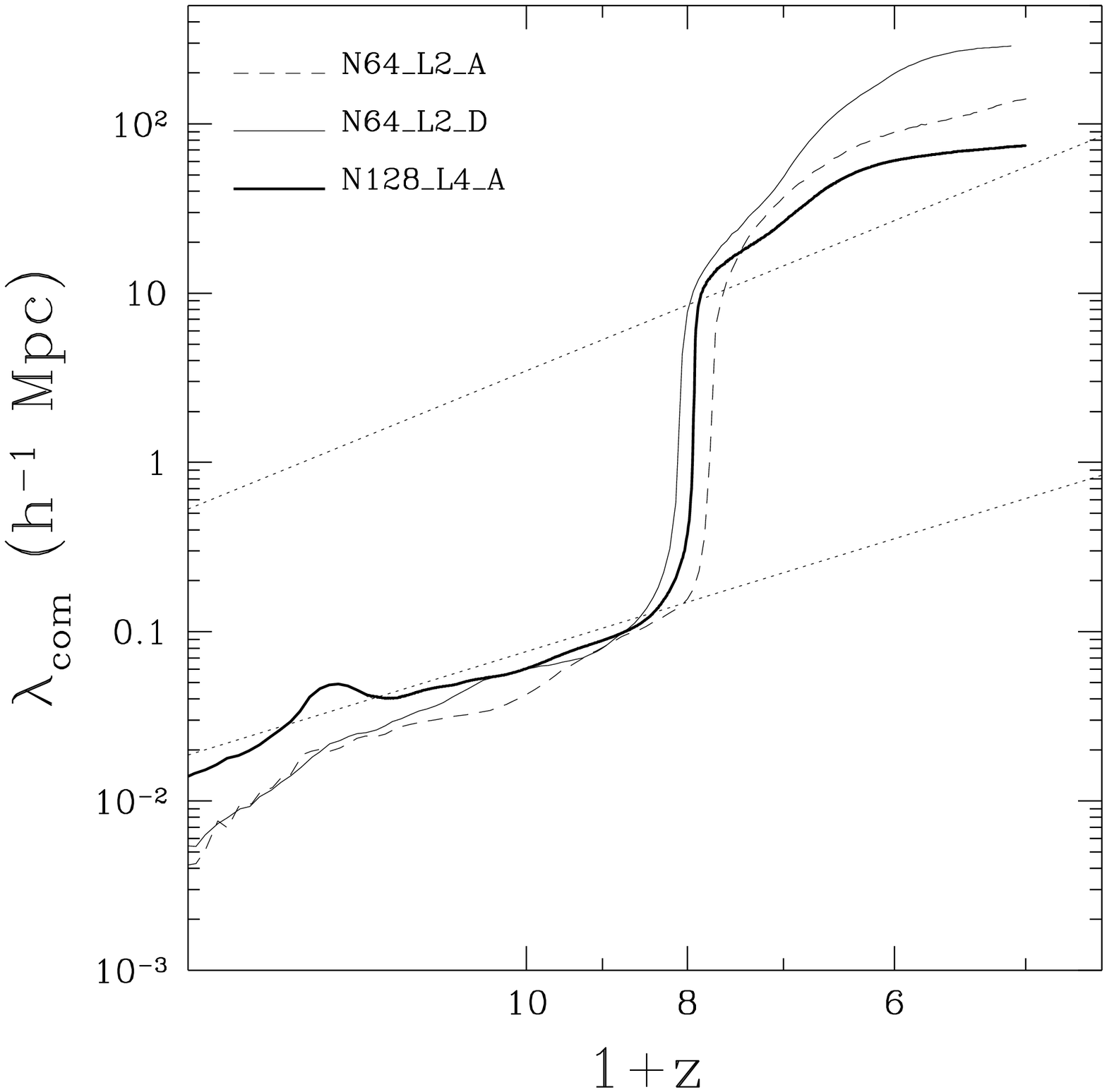}{\figdir/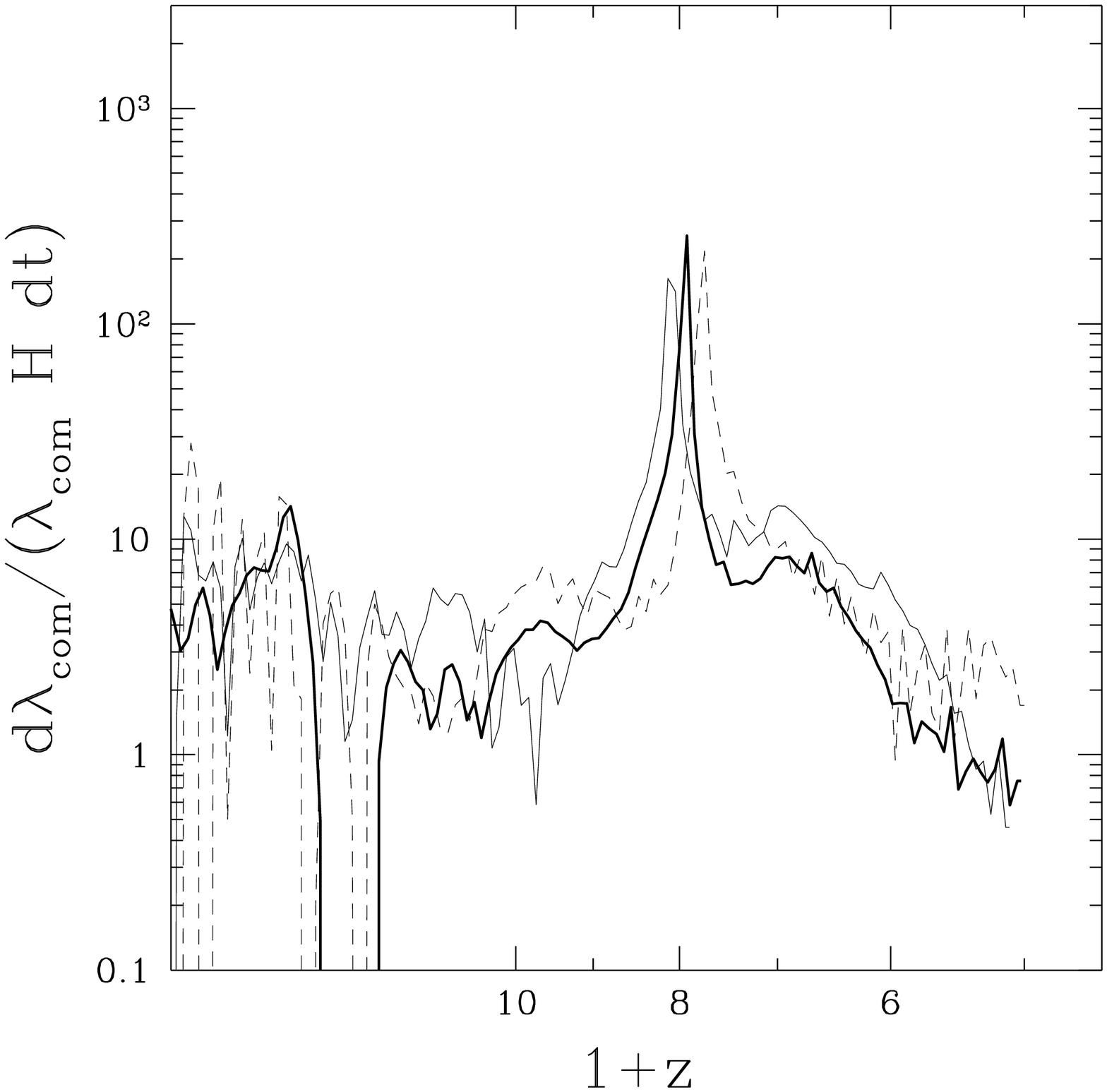}
\caption{\label{figMP}\capMP}
\end{figure}
}
Let me now focus on details of the process that is overviewed in the
previous subsection. Figure \ref{figMP} shows the evolution of the mean free
path and its time derivative for the production run N128\_L4\_A and two
smaller runs (the latter two to demonstrate the level of numerical
convergence). The epoch of overlap is clearly distinguished by a
sharp rise in the mean free path. The rate of this rise, some 100 times
faster than the expansion of the universe, shows no trend of diminishing
with the increase of the box size (the bold line versus thin ones), and
thus is unlikely to be an artifact of a finite simulation box.

Can this behavior be understood in simple terms? Let me consider
a simple model for the universe: each object has a $1/r^2$ density
profile around it until those profiles overlap with neighboring
objects. A source, placed at the center of an object produces 
the photoionization rate that falls off as $1/r^2$ in the optically
thin regime,
$$
	\Gamma(r) = \Gamma_0 {r_0^2\over r^2}.
$$
Given the density profile with a core radius $r_0$,
$$
	n(r) = n_0 {r_0^2\over r_0^2+r^2},
$$
the neutral hydrogen fraction at $r\gg r_0$ is constant,
$$
	x(r) = x_0 = { R n(r)\over \Gamma(r) } = { R n_0\over \Gamma_0 }.
$$
The mean free path then is given by the following expression,
$$
	\lambda = {c\over\bar{k}},
$$
where $\bar{k}$ is the photoionization rate averaged value of the
absorption coefficient (I omit here frequency dependence for simplicity):
\begin{equation}
	\bar{k} = {\langle k\Gamma\rangle\over\langle\Gamma\rangle}
	\label{eqkbar}
\end{equation}
(see Gnedin \& Ostriker 1997, eqn.\ [A6]), and $\langle\Gamma\rangle$ is the
photoionization rate spatially averaged over a sphere of radius $R$,
$$
	\langle\Gamma\rangle = {3\over R^3}
	\int_0^R \Gamma(r) r^2dr = 3\Gamma_0 
	{r_0^2\over R^2}.
$$
Using the expression for the density, for the mean free
path I obtain:
\begin{equation}
	\lambda = {2\over \pi \sigma x_0 n_0} {R\over r_0},
	\label{mfp}
\end{equation}
where $\sigma$ is the ionization cross section.
If, instead of $\bar{k}$, I used a spatially averaged absorption coefficient
$\langle k\rangle$, which would be appropriate if the ionizing background was
homogeneous, I would get a different result:
\begin{equation}
	\hat\lambda \equiv {c\over\langle k\rangle} = {1\over 3 \sigma x_0 
	n_0} {R^2\over r_0^2}.
	\label{mfphat}
\end{equation}
Reasonable values for $n_0$ and $r_0$ would be the
virial density, $n_0=200\bar{n}_b=4\times10^{-5}(1+z)^3
\dim{cm}^{-3}$, where $\bar{n}_b$ is the average baryon density, and 
the virial radius, $r_0=31(1+z)^{-1}\dim{kpc}$ for a $10^9$ solar mass
object. Obviously, $R$ has to be about the size of the $\HII$ region around
the source, which from Fig.\ \ref{figIM} can be roughly approximated as
$R=10(1+z)^{-2}\dim{Mpc}$ at $z\sim9$. I also assume $x_0\sim10^{-2}$, 
which is justified below. 

The two dotted lines in Fig.\ \ref{figMP}a now show $\lambda$ (the
lower line)
and $\hat\lambda$ (the upper line). 
Before the overlap, the radiation field
is highly nonuniform, and the mean free path is short, as the majority of
photons are absorbed within one $\HII$ region. 
The lower dotted line ($\lambda$
from eqn.\ [\ref{mfp}]) provides in this regime a remarkably good fit
to the simulation results despite the highly simplistic nature of the
above derivation. Fig.\ \ref{figIM}b-d illustrate this effect. The
$\HII$ region located in the middle right of the slice
contains the high density neutral
filaments, which are slowly being eaten away by the ionization front,
which moves much slower across high density filaments than over the low
density void. The absorption in these high density regions shortens the
mean free path compared to the size of the $\HII$ region in the low density
medium. This is consistent with the simplistic picture above which
assumes a spherically symmetric density distribution, which in reality is
an average of the high density filaments (with the density run shallower than
$1/r^2$) and the low density voids (with the density run steeper than
$1/r^2$).

\def\capJD{
The joint mass-weighted 
distribution of the gas density and the local photoionization rate $\Gamma$
in the production run N128\_L4\_A at four different redshifts. The straight 
bold line in the upper left panel marks a
$\Gamma=\langle\Gamma\rangle(1+\delta)$ law.
}
\placefig{
\begin{figure}
\insertfigure{\figdir/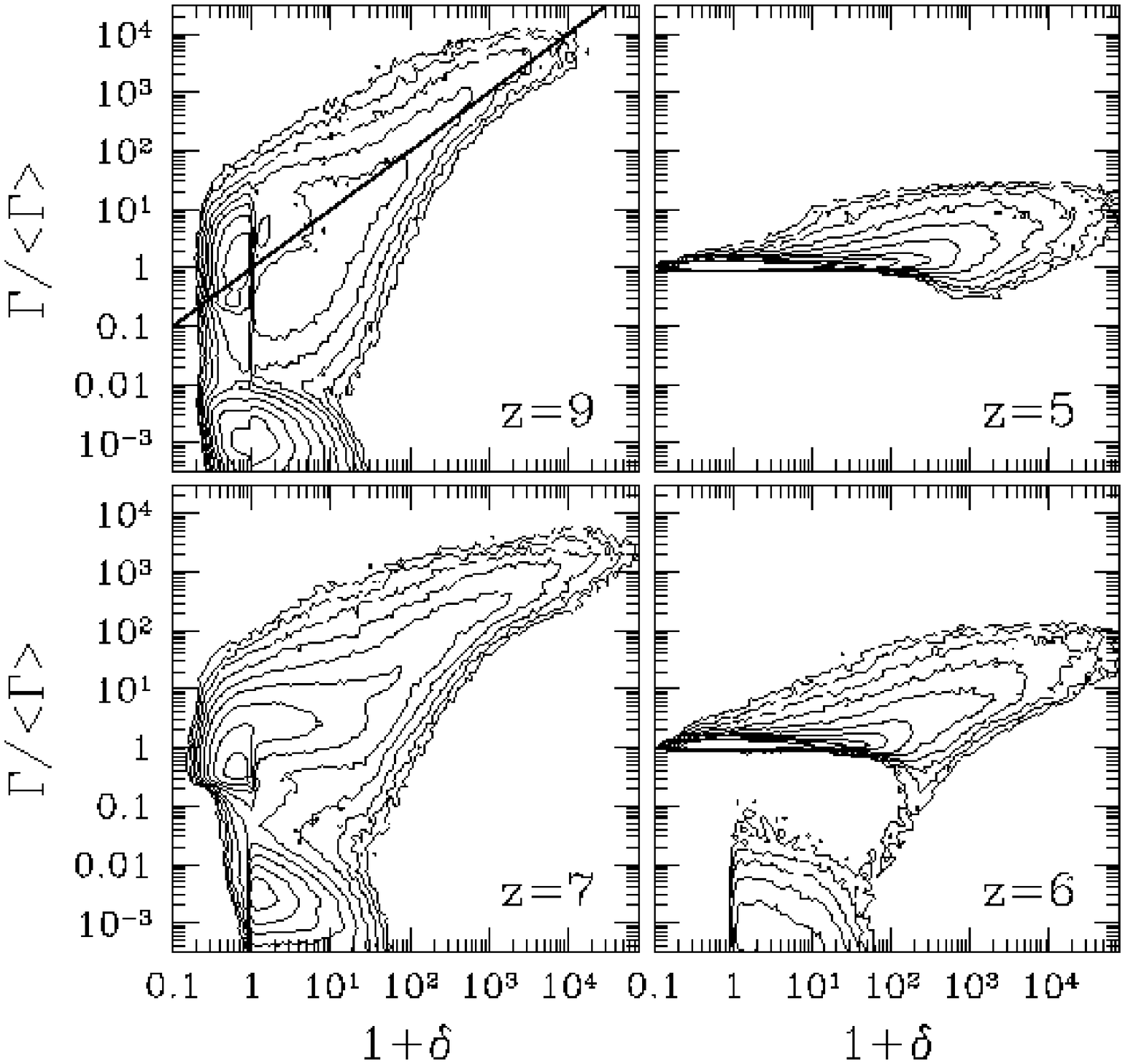}
\caption{\label{figJD}\capJD}
\end{figure}
}
After the overlap, the ionizing
radiation is more (albeit non perfectly) homogeneous, and the mean free
path increases by almost two orders of magnitude and becomes comparable to the
estimate $\hat\lambda$. Thus, the effect of the overlap being ``fast'' 
is entirely due to the
change in the relative distribution of the ionizing radiation and
the gas density, and thus, in essence, is a topological effect.

\def\capXD{
The joint mass-weighted 
distribution of the gas density and neutral hydrogen fraction
in the production run N128\_L4\_A at four different redshifts.
The excessive sharpness of the features at $\delta\approx0$ is the defect 
of the
radiative transfer approximation, in reality there should be a more gradual
transition from the underdense to the overdense regions.
}
\placefig{
\begin{figure}
\insertfigure{\figdir/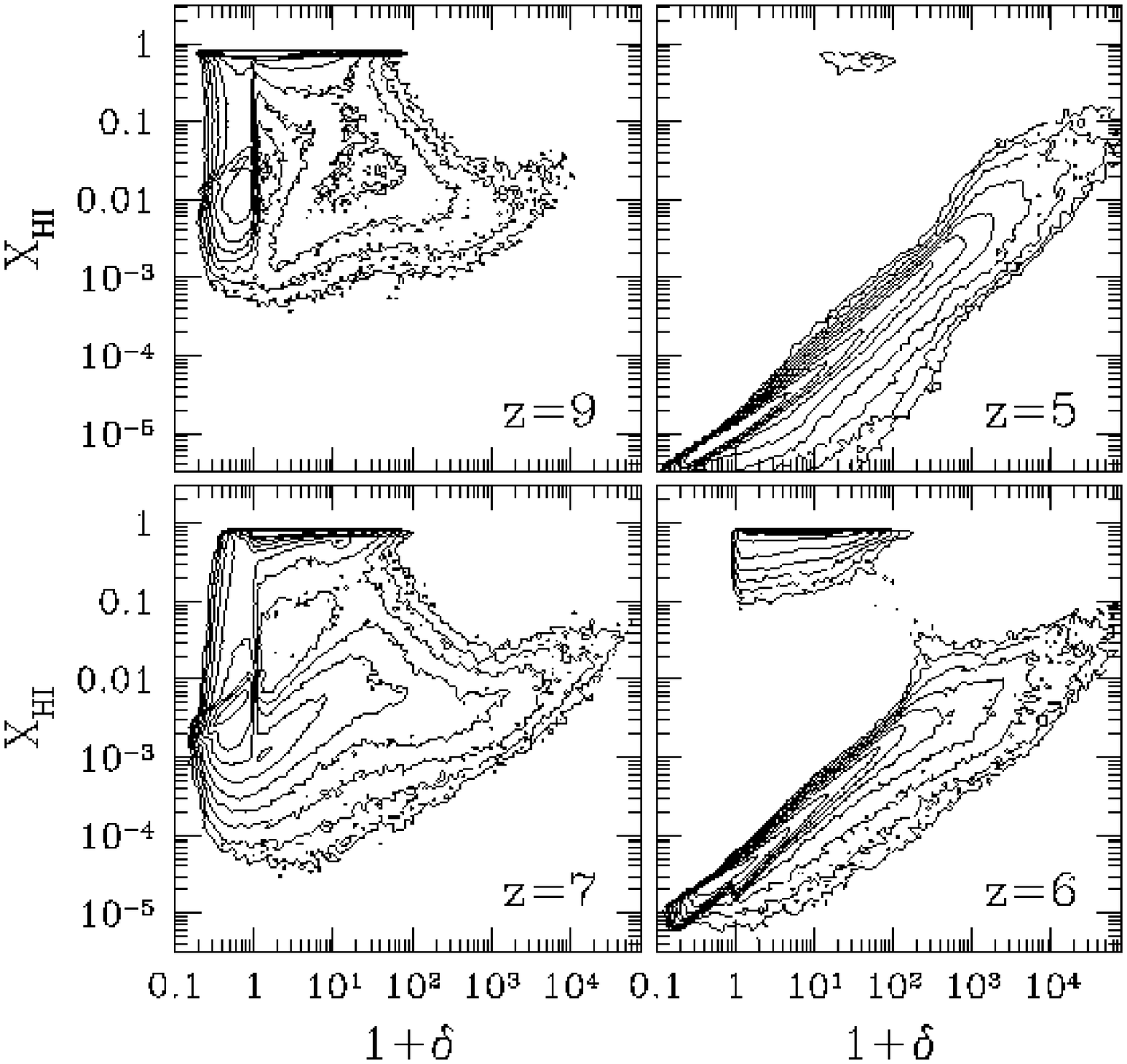}
\caption{\label{figXD}\capXD}
\end{figure}
}
This is further illustrated by Figure \ref{figJD}, which shows on four
panels the joint mass-weighted distribution of the gas density and the
photoionization rate at four different redshift. At $z=9$ the photoionization
rate in the regions that are indeed photoionized (only those count towards
the mean free path, since no photons are absorbed where there are no 
photons) is proportional to the gas density, with a large scatter reflecting
the fact that different objects have different luminosities (in terms of
the simplistic model above, $\Gamma_0$ is different in different objects).
As the $\HII$ regions start to overlap ($z=7$), regions with more or less
uniform photoionization rate starts to appear at low densities. Finally,
in the post-overlap stage, a large fraction of the total volume has a
quite uniform photoionization rate, which still varies significantly in
the high density regions ($z=6$ and $z=5$). I will address the question of
the inhomogeneity in the photoionization rate at low redshifts below.

Figure \ref{figXD} illustrates the process of reionization from yet
another point of view. It presents
the mass-weighted joint
distribution of the gas density and neutral hydrogen fraction shown 
at the same four redshifts $z=9$, 7, 6, and 5.
In the pre-overlap stage the neutral hydrogen fraction is roughly
independent of density in the photoionized regions, again with a
large spread due to different luminosities of different objects and
with the median value of about $x_{\HI}\sim10^{-2}$, which is a
justification for adopting this value in the simplistic estimate
presented above.
As time
progresses and the $\HI$ regions start to overlap, a regime 
$x_{\HI}\propto(1+\delta)$ starts to develop in the lower density
gas, which is just a reflection
of the fact that the photoionization rate is becoming independent of
the density in those regions, whereas the high density gas still maintains
the neutral fraction that is roughly independent of the density.
After the overlap, most of the gas sits at the ionization equilibrium
in a tight relation with the density, which becomes somewhat more spread in
the highest density regions which correspond to the inner parts of
individual objects. 
However, a considerable fraction of the high density gas
remains in the neutral state, in accordance with the semi-analytical
picture of Miralda-Escud\'{e} et al.\ (1999). As the production of the
ionizing photons continues, this gas is slowly being ionized, mostly
the lower density regions first.

One of the outcomes of the main assumption of this paper - that the sources
of ionization are proto-galaxies rather than quasars - is that each halo
harbors a source (except, perhaps, very low mass ones, that are not
well resolved in my simulation), and therefore the halos
are also ionized by local sources. The resolution of my simulation is
however insufficient to resolve the interstellar medium, and proto-galaxies
will contain giant molecular clouds which will be responsible for the damped
Lyman-alpha absorption.

\subsection{What does it?}

\def\capLA{
The star formation rate versus the stellar mass ({\it a\/}) and
the virial radius versus the Str\"omgen radius ({\it b\/}) 
for all bound objects in the production run N128\_L4\_A.
}
\placefig{
\begin{figure}
\epsscale{0.65}
\inserttwofigures{\figdir/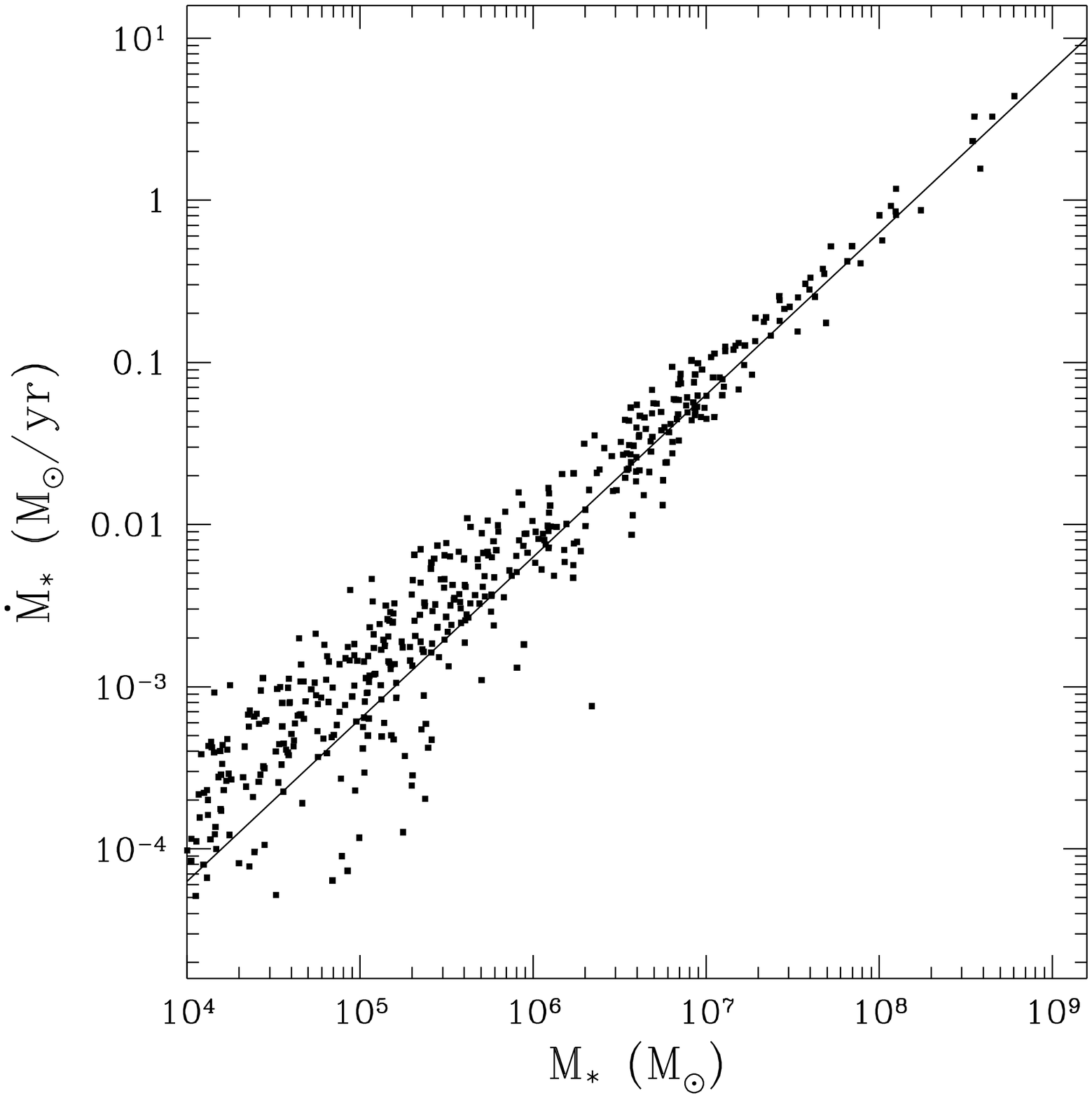}{\figdir/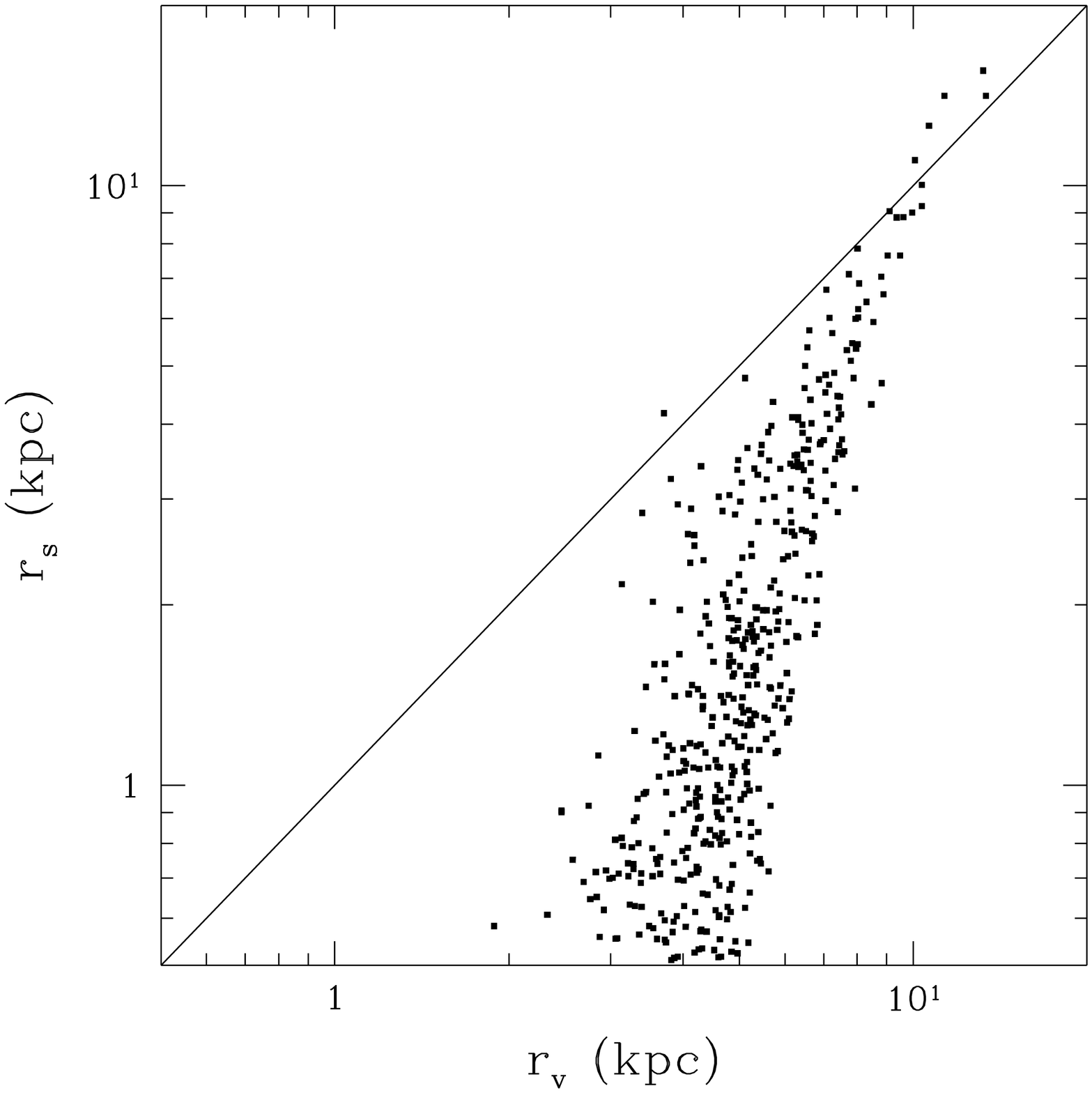}
\caption{\label{figLA}\capLA}
\end{figure}
}
The understanding of the process of reionization would not be complete
without identifying which sources are indeed responsible for the bulk
of ionizing photons. To this end, I have identified all bound objects
in the production run N128\_L4\_A at $z=7$ (the epoch of overlap)
using Bertchinger \&
Gelb's (1991) DENMAX algorithm. Figure \ref{figLA}a shows the star formation
rate versus the stellar mass for all objects. As can be seen,
there exists a good correlation between the star formation rate and the
mass of the object, 
$$
	\dot{M}_* \approx {M_*\over1.6\times10^8\dim{M}_\odot} 
	{\dim{M}_\odot\over \dim{yr}},
$$
reflecting the fact that all objects found in the
simulation form stars with the rapidly increasing rate (Fig.\ \ref{figSF}).

However, not all objects contribute equally toward the reionization of the
low density IGM, but only those whose ionization fronts are able to leave
their own halo. While the detailed analysis of the evolution of each
individual ionization front is beyond the scope of this paper, a simple
picture can be introduced that captures the main qualitative features of
the simulation.

If each individual object can be approximated as a homogeneous sphere with
the radius equal the virial radius of the object, and the mean overdensity
of 200, than an object will contribute towards the reionization of the
low density IGM if the virial radius $r_v$
is smaller than the Str\"omgen
radius $r_s$ of the ionization front. Figure \ref{figLA}b now shows the
the Str\"omgen
radius versus the virial radius for all the bound resolved objects in the
production run. One can see that the majority of objects have their
Str\"omgen spheres well inside the virial radius, and only a handful of
objects contribute to the ionizing flux that is capable of ionizing the
low density IGM. 

\def\capLF{
The stellar mass function ({\it solid bold line\/}) and the star
formation rate function ({\it dashed line\/}) at $z=7$.
The dotted line shows the modified star formation rate function
which includes only photons capable of ionizing the low density IGM.
The thin solid line is a Schecter function fit to the stellar mass
function with $\alpha=-0.5$ and $M_\star=3\times10^8\dim{M}_\odot$.
}
\placefig{
\begin{figure}
\insertfigure{\figdir/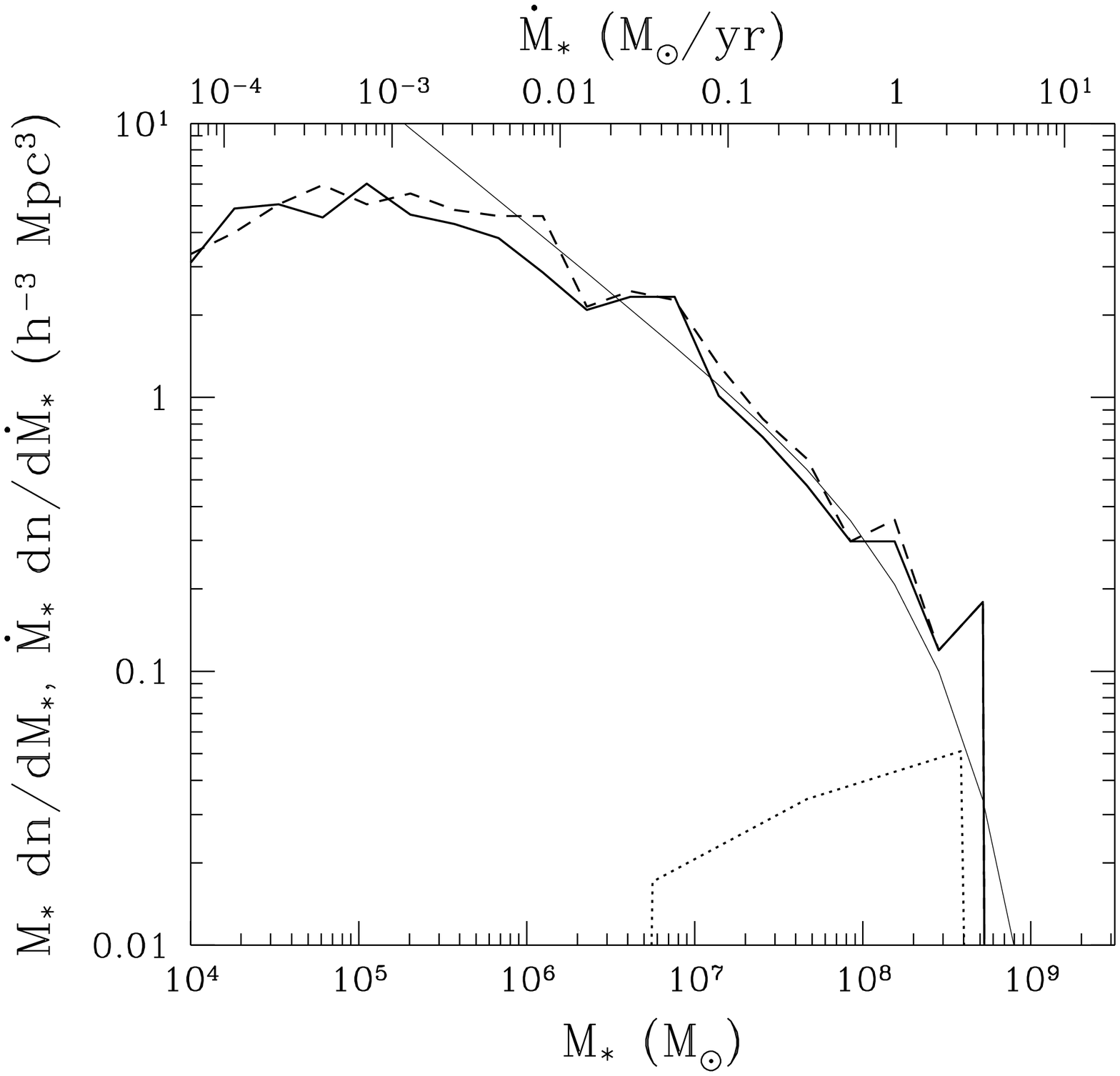}
\caption{\label{figLF}\capLF}
\end{figure}
}
This is not surprising, given what is seen in Fig.\ \ref{figIM}, and the
fact that I have not been able to establish numerical convergence -
if only a few objects in my simulation box are responsible for
reionization, then the $4h^{-1}\dim{Mpc}$
box size is too small to be a representative volume of the universe, and
a larger box size is needed to achieve the full numerical convergence.

Figure \ref{figLF} serves to illustrate this further. It shows the
stellar mass function and the star formation rate function for the
bound objects from my simulation, as well as the Schecter function fit
to the stellar mass function (the fit deviates at low masses due to
photoionization feedback). In addition, I plot with the dotted line
the modified star formation rate function
which includes only photons capable of ionizing the low density IGM,
defined as
$$
	\dot{M}_I {dn\over d\dot{M}_I},
$$
where
$$
	\dot{M}_I = \max(0,(r_s/r_v)^3-1)\dot{M}_*
$$
is proportional to the number of photons that escape into the IGM.
As can be seen from Fig.\ \ref{figLF}, it is objects with luminosities
in excess of $0.1L_\star$ that are responsible for reionization, with
$L_\star$ and more luminous objects making the main contribution.

\subsection{Comparison with the semi-analytical models}

\def\capDE{
({\it a\/})
Comparison of the mean volume emissivity $\epsilon$ ({\it solid line\/}), 
measured in terms
of the number of ionizing photons emitted for each atom in the universe per 
Hubble time, and the global recombination rate $R$ ({\it dashed line\/}),
measured in terms of the mean number of recombinations per Hubble time per
baryon, as introduced in equation (2) of Miralda-Escud\'{e} et al.\ (1999).
({\it b\/}) The time evolution of the quantity $\Delta_i$ as defined in
equation (1) of Miralda-Escud\'{e} et al.\ (1999).
}
\placefig{
\begin{figure}
\epsscale{0.65}
\inserttwofigures{\figdir/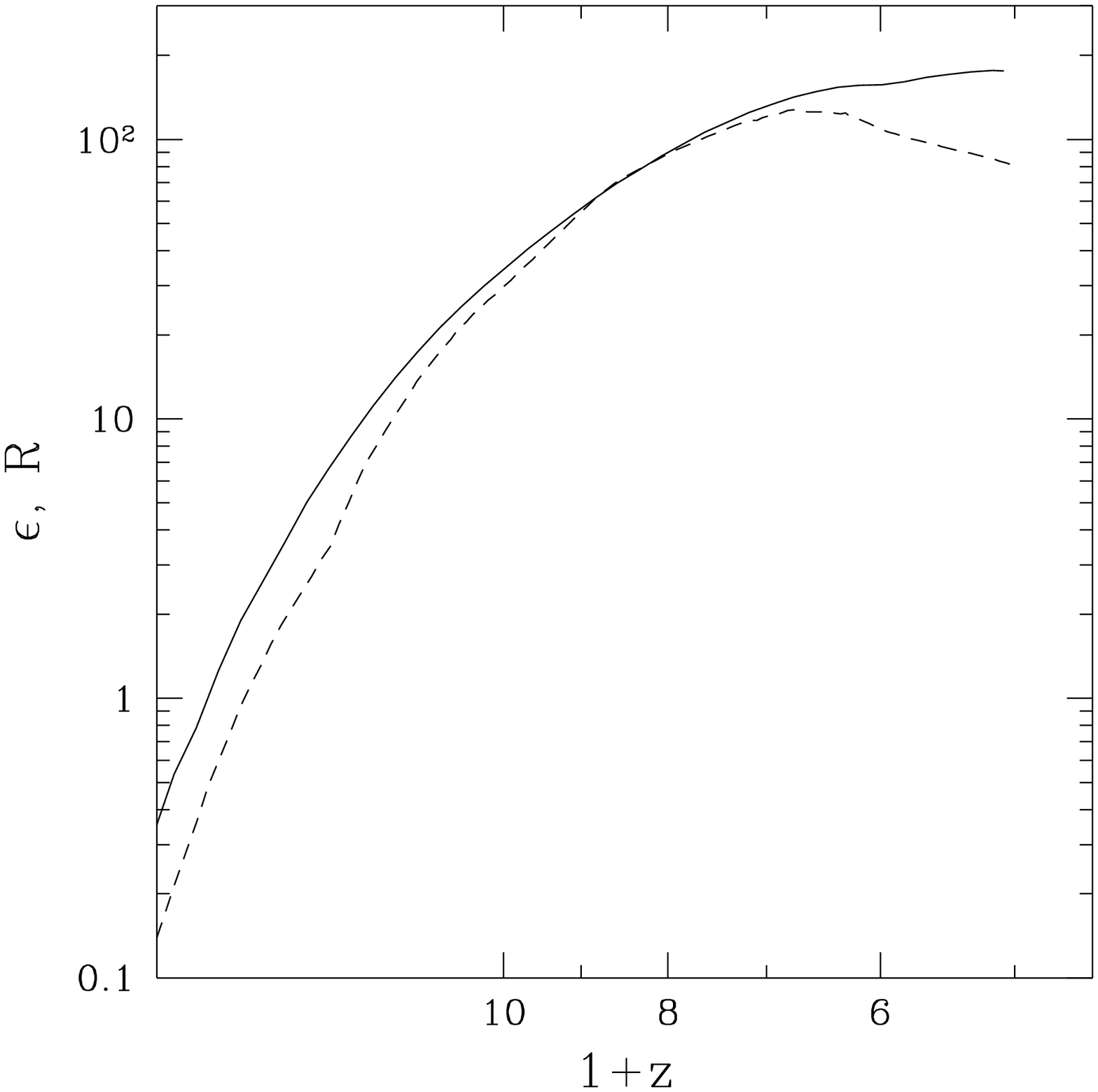}{\figdir/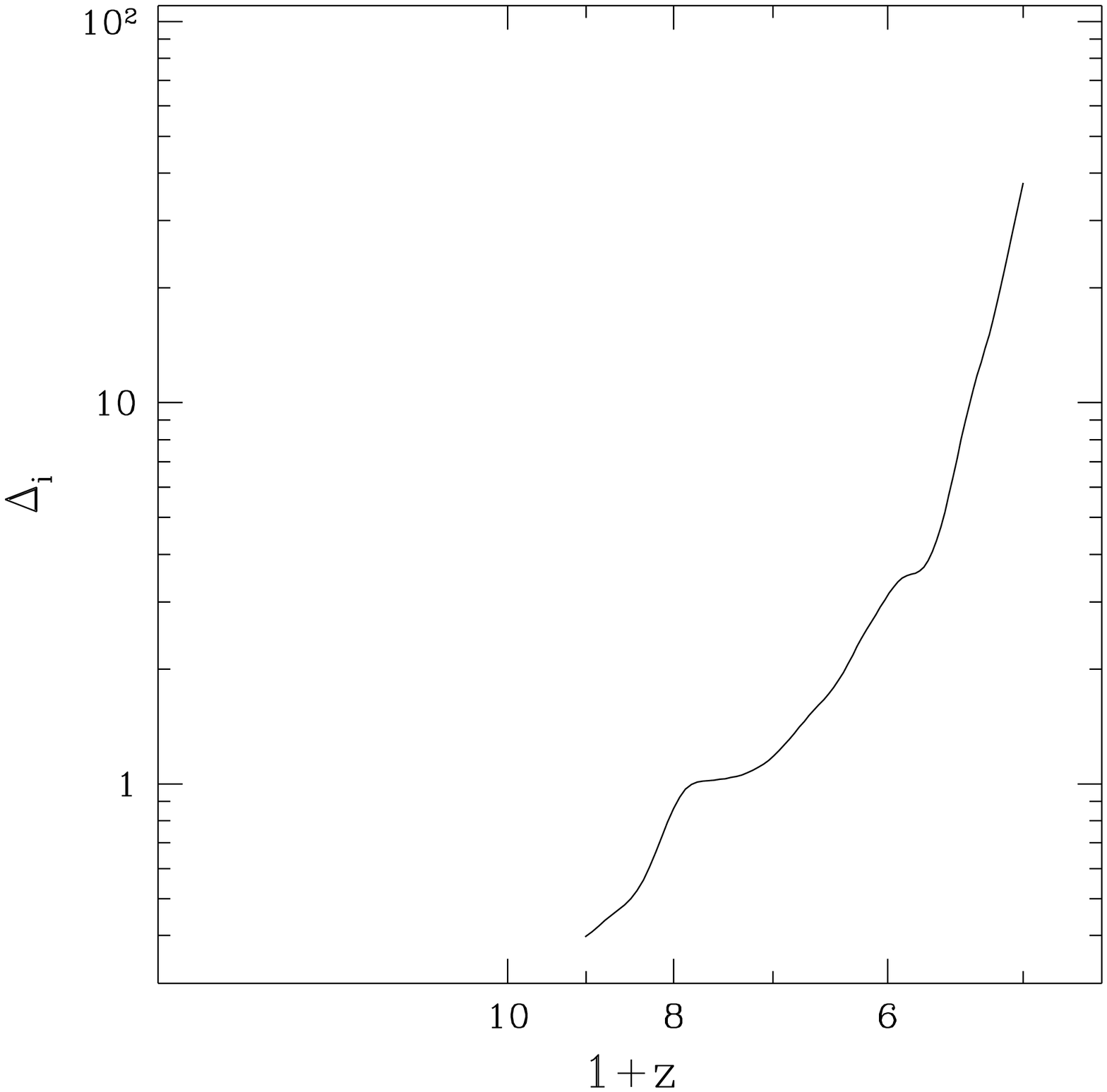}
\caption{\label{figDE}\capDE}
\end{figure}
}
I have mentioned already that my simulations agree with the qualitative 
predictions of Miralda-Escud\'{e} et al.\ (1999) quite well 
in the post-overlap stage.
Figure \ref{figDE} serves to illustrate the agreement even further. 
Fig.\ \ref{figDE}a shows a comparison between the mean volume emissivity
$\epsilon$ and the global recombination rate $R$, 
as introduced in equation (2) of
Miralda-Escud\'{e} et al.\ (1999). In accordance with their prediction,
$R \approx\epsilon $ just after the overlap, and continues to be close to
$\epsilon$ and somewhat lower when other terms become important.

Fig.\ \ref{figDE}b shows the evolution of the quantity $\Delta_i$
defined in equation (1) of Miralda-Escud\'{e} et al. (1999). Physically,
$\Delta_i$ measures the cosmic density (in units of the mean density)
above which the gas is mostly neutral, and below which the gas is mostly
ionized. While Fig.\ \ref{figXD} demonstrates that at high density
the ionization state of the gas is not related to the density in any
simple way, but also depends on whether a given fluid element is close
or far from the nearest source, it is still possible to define $\Delta_i$
after the overlap as the limiting density above which 95\% of the neutral
gas lies. This quantity is plotted in Fig.\ \ref{figDE}b. While specific
numerical values for $\Delta_i$ are significantly lower than those quoted in
Miralda-Escud\'{e} et al. (1999), because there is also the high density
ionized gas responsible for a large number of recombinations, the general
trend of $\Delta_i$ increasing  with time is clearly observed.

\def\capNG{
{\it Top panel\/}: the mean number of ionizing photons emitted per baryon 
as a function of
redshift for the production run N128\_L4\_A ({\it bold line\/})
and a smaller run N64\_L2\_D ({\it thin line line\/}).
{\it Bottom panel\/}: the mass- ({\it dotted lines\/}) and volume-averaged
({\it solid lines\/}) neutral hydrogen fractions for the production run 
N128\_L4\_A ({\it bold lines\/})
and a smaller run N64\_L2\_D ({\it thin solid lines\/}).
}
\placefig{
\begin{figure}
\insertfigure{\figdir/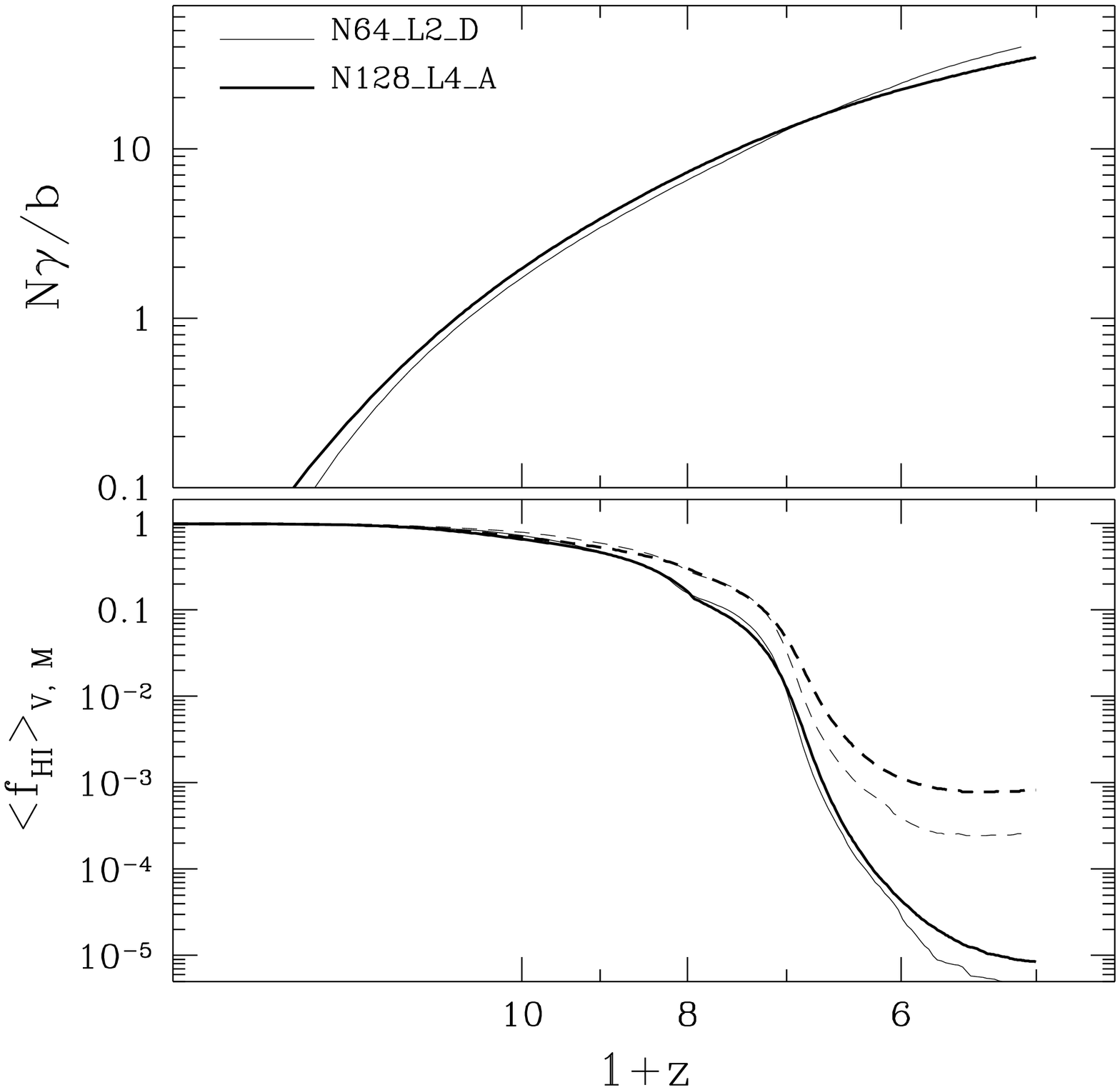}
\caption{\label{figNG}\capNG}
\end{figure}
}
Figure \ref{figNG} shows the evolution of the mean number of ionized photons
per baryon $N_{\gamma/b}$ on top and the neutral hydrogen fraction at 
the bottom for the
production run and for the corresponding small run. One can see that by the
time of overlap $z=7$, about 10 ionizing photons were emitted per baryon,
which may be considered a contradiction with Miralda-Escud\'{e} et al. (1999).
I must however reemphasize here that the total number of photons emitted
depends on the radiation efficiency $\epsilon_{UV}$, which in turn is
highly resolution dependent. Therefore the quantity $N_{\gamma/b}$ is also
resolution dependent and has no physical meaning taken at a face value.
The quantity used by Miralda-Escud\'{e} et al. (1999) is however the
number of ionizing photons per baryon escaped into the IGM from the immediate
neighborhood of the ionizing sources,\footnote{This quantity is physically
intuitive, but is difficult to define in a mathematically rigorous fashion.}
 and is therefore considerably smaller
than the number directly taken from the simulation. The latter quantity cannot
be much less than 1, and since it has to be considerably smaller than 10,
the best estimate one can make is that it is of order of a few, again in
agreement with Miralda-Escud\'{e} et al.\ (1999).

A large number of workers has studied reionization within the framework
of the semi-analytical modeling using the clumping factor approach (see
the references in the Introduction).
Their main conclusion is that reionization is fast (because the recombination
time is short), and that many ionizing photons per baryon are required 
to ionize the universe (again because of the same reason). It is therefore
useful to compare this approach to the simulations and to
Miralda-Escud\'{e} et al.\ (1999). The latter comparison is especially
instructive since there appears to be a complete disagreement between the
two semi-analytical approaches: if Miralda-Escud\'{e} et al.\ (1999) claim
that reionization has to be gradual, the clumping factor approach predicts
a fast reionization; if Miralda-Escud\'{e} et al.\ (1999) show that just a
few (perhaps as low as one) ionizing photons per baryon are needed for
the overlap, the clumping factor approach requires that this number be large.

\def\capCF{
({\it a\/}) The mean global recombination rate in units of the 
recombination rate of the fully ionized homogeneous universe
({\it solid line\/}), the density clumping factor ({\it dotted line\/}),
and the ionized hydrogen clumping factor ({\it dashed line\/}) as a function
of redshift. 
({\it b\/}) The porosity of the IGM
as a function of redshift for the approximation
based on equation (23) of Madau et al.\ (1999) ({\it dashed line\/}) and
the full calculation based on their equation (21) ({\it solid line\/}).
}
\placefig{
\begin{figure}
\epsscale{0.65}
\inserttwofigures{\figdir/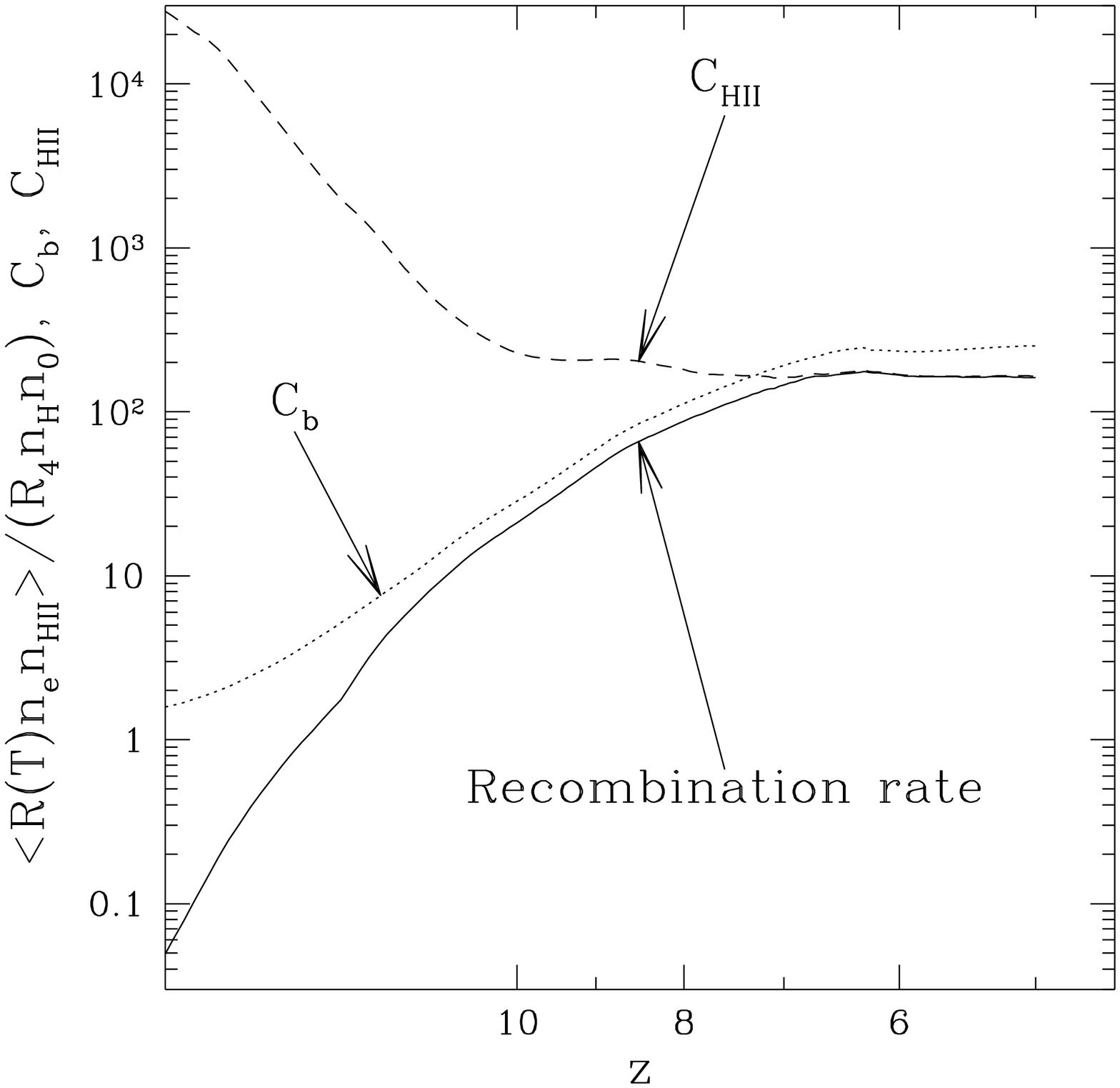}{\figdir/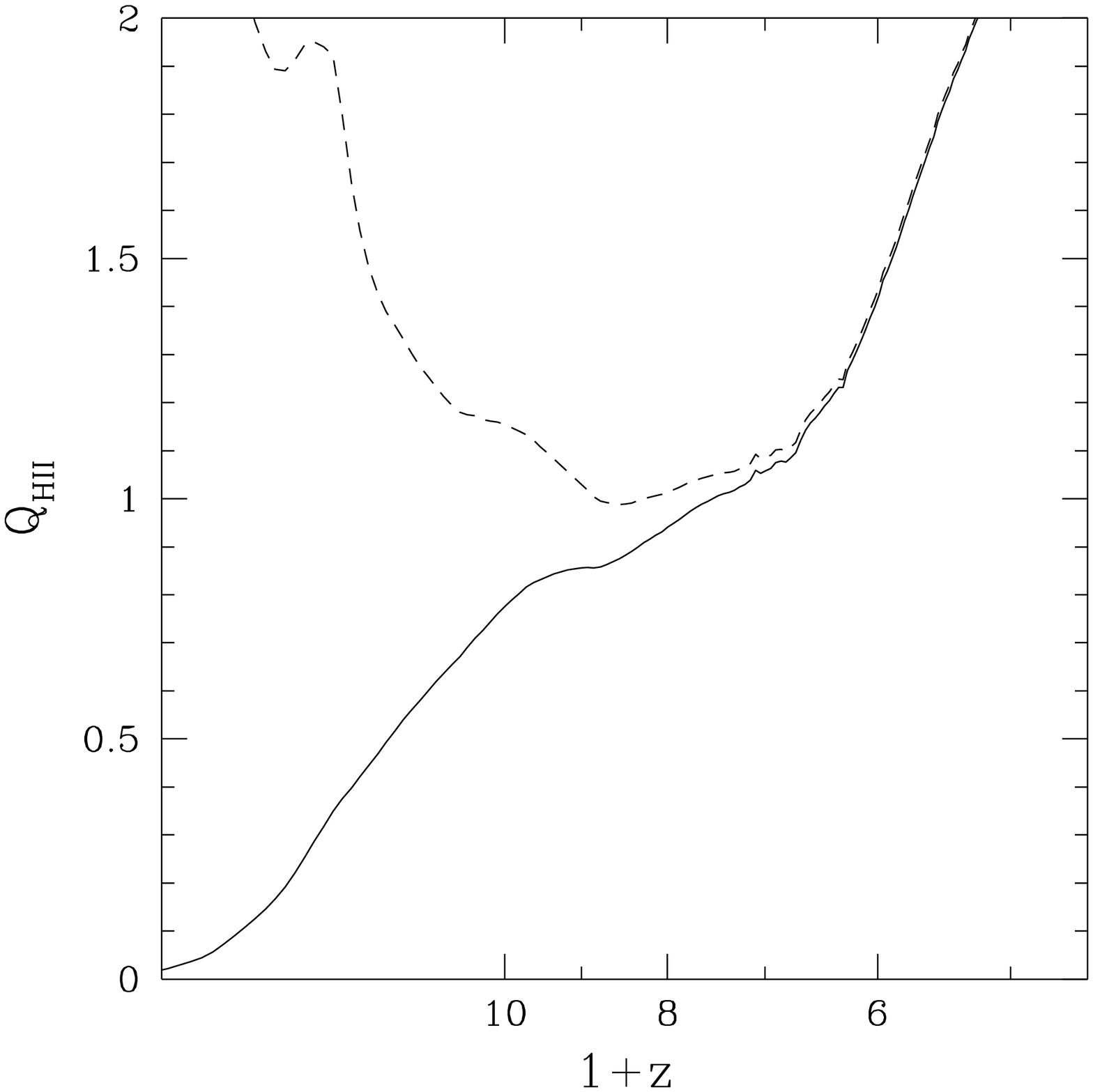}
\caption{\label{figCF}\capCF}
\end{figure}
}
I now show in Figure \ref{figCF}a the evolution of the volume averaged
recombination
rate $\langle R(T)n_en_\HII\rangle$ in units of the recombination rate
for the fully ionized homogeneous universe with $T=10^4\dim{K}$
($R_4\equiv R(10^4\dim{K})$, $n_{\rm H}$ is the total number density of
atomic and ionized hydrogen, and $n_0=n_{\rm H}+n_{\rm He}$, thus assuming
that helium is only singly ionized). Also shown the density clumping factor
$$
	C_b \equiv {\langle\rho^2\rangle\over\langle\rho\rangle^2},
$$
and the ionized hydrogen clumping factor
$$
	C_\HII \equiv {\langle R(T)n_en_\HII\rangle\over
	R_4 \langle n_e\rangle\langle n_\HII\rangle},
$$
which is simply related to the volume averaged recombination rate,
\begin{equation}
	\langle R(T)n_en_\HII\rangle = R_4 \bar{n}_e
	\bar{n}_\HII C_\HII,
	\label{rrvscf}
\end{equation}
where bar means volume average.

There are few things to note. First, the recombination rate increases fast
with time, and at the epoch of overlap already approaches 100 times that
of the homogeneous universe. The density clumping factor is only slightly
higher than the recombination rate, and the ionized hydrogen clumping 
factor begins its evolution with very high values, and approaches the
recombination rate at later times, as can be expected from equation
(\ref{rrvscf}). This behavior is quite different from what was predicted
in Gnedin \& Ostriker (1997), and the reason for the disagreement is quite
clear: since the ionization fronts originate in the highest density
regions, the first $\HII$ regions are extremely overdense, which results
in very large values for the clumping factor. However, since these
first $\HII$ regions occupy only a small fraction of the volume, the
recombination rate is not large.

Thus, the simulations confirm the predictions of the clumping factor approach:
the clumping factor is indeed large at the time of overlap. 
To do a more detailed comparison, I picked up a Madau et al.\ (1999) paper
as an example of semi-analytical modeling using this type of approach.
There have been recently a series of detailed papers on using this
method (Haiman \& Loeb 1997, 1998;
Valageas \& Silk 1999; Chiu \& Ostriker 1999), but Madau et al.\ (1999)
paper presents the clumping factor approach in its essence, in the
simplest possible flavor, and therefore is easy to reproduce.

Figure \ref{figCF}b now shows the most important quantity in the
semi-analytical modeling: the porosity of the IGM $Q_\HII$
as a function of time.
The dashed line shows the approximate expression from equation (23) of
Madau et al.\ (1999) paper, and the solid line reflects a more
accurate calculation based on equation (21) of Madau et al.\ (1999) paper
with the comoving emissivity extracted from the simulation. Again, the
main conclusion of Madau et al.\ (1999) is confirmed, the overlap
occurs when $Q_\HII\approx1$.

I have just demonstrated, that my simulation agrees (at least on a
semi-qualitative level) with semi-analytical models of 
Miralda-Escud\'{e} et al.\ (1999) and Madau et al.\ (1999). Then
how about the disagreement between the two? Clearly, if both agree
with the same simulation, they also agree with each other. Let me
know discuss the ``differences'' between the two semi-analytical
approaches.

First of all, if one looks carefully, one can notice that 
equation (23) of Madau et al.\ (1999) is indeed the ratio
of $\epsilon$ to $R$ using Miralda-Escud\'{e} et al.\ (1999) 
notation, and the Miralda-Escud\'{e} et al.\ (1999) equation (3)
$\epsilon\approx R$ is indeed equation (24) of Madau et al.\ (1999).
In other words, the dashed line in Fig.\ \ref{figCF}b is a ratio of
the two curves in Fig.\ \ref{figDE}a.
There is therefore no principal difference between the two approaches,
but the advantage of Miralda-Escud\'{e} et al.\ (1999) approach is that
they propose a model for the clumping factor based on the realistic 
density distribution function, 
rather than make an ad hoc assumption about the time evolution of
the clumping factor. The main disadvantage of their model is that
it is not necessarily accurate quantitatively.\footnote{Miralda-Escud\'{e} 
et al.\ (1999) model will work much better quantitatively for the case when
the universe was reionized by a few bright quasars. In this case the
amount of gas in the vicinity of sources is much smaller than in the
case of stellar sources of reionization, when there are sources in
a good fraction of all high density regions.}

There however still remains a question of the number of ionizing photons
per baryon required to ionized the universe. It cannot be simultaneously
one (Miralda-Escud\'{e} et al.\ 1999) and many (Madau et al.\ 1999),
can it?

The irony of the situation is that it indeed can. The difference between
these two approaches is mostly terminological, and it can be best understood
if one considers the number of ionizing photons per baryon as a function of
scale, $N_{\gamma/b}(R)$ (this quantity cannot be rigorously defined
mathematically, but has an intuitively clear physical sense).
If Madau et al.\ (1999) count all photons emitted by all sources
(in our case stars), i.e.\ they consider a quantity $N_{\gamma/b}(0)$
(here zero actually means the stellar surface), then 
Miralda-Escud\'{e} et al.\ (1999) use $N_{\gamma/b}(\mbox{IGM})$
where the symbol $\mbox{IGM}$ stands for the photons that escape local 
absorption,
i.e.\ absorption in the gas bound to a given source, and
corresponds to a range of scales of the order of $0.1-1\dim{Mpc}$. 
The top panel of Fig.\ \ref{figNG} gives yet another quantity,
$N_{\gamma/b}(R_{\mbox{min}})$, where $R_{\mbox{min}}$ is the resolution of the
simulation, and in my case it is 
$$
	R_{\mbox{min}} = {1\over1+z}h^{-1}\dim{kpc}
$$
in physical units.
It is now easy to bridge a difference between Madau et al.\ (1999)
and Miralda-Escud\'{e} et al.\ (1999). The number $N_{\gamma/b}(0)$ is
perhaps a factor of a few larger than $N_{\gamma/b}(R_{\mbox{min}})
\approx10$ at
$z=7$, which
gives a value for $1/N_{\gamma/b}(0)$ of the order of a few percent, 
in complete
agreement with Madau et al.\ (1999). On the other hand, as has been
discussed above, $N_{\gamma/b}(\mbox{IGM})$ should be considerably smaller
than $N_{\gamma/b}(R_{\mbox{min}})$, which leaves us with the conclusion of
Miralda-Escud\'{e} et al.\ (1999) that $N_{\gamma/b}(\mbox{IGM})$ is of the
order unity.

There remains one final sticking point: a question of whether reionization
is ``fast'' or ``slow''. But I have already mentioned in \S\ref{sec:rg} that
it is to a some degree a question of terminology as well. 
The whole process of reionization
consists of three different stages: the ``slow'' (of the order of the Hubble 
time) pre-overlap, the ``fast'' (of the order of one tenth of the Hubble 
time) overlap, and the ``slow'' post-overlap. Miralda-Escud\'{e} et al.\ 
(1999) focused on the post-overlap stage, but considered the whole process
of reionization, and therefore concluded that reionization was slow.
\footnote{There may however still remain a difference, since 
Miralda-Escud\'{e} et al.\ (1999) also claimed that the overlap is
``gradual''. It is not clear however without further elaboration of
the Miralda-Escud\'{e} et al.\ (1999) model whether this ``gradual''
overlap is actually ``fast'' enough to be compatible with other
semi-analytical approaches and with this work
(after all, the ionizing intensity has to
increase by two to three orders of magnitude over a Hubble time or so),
or whether there is
a genuine disagreement between Miralda-Escud\'{e} et al.\ (1999) and
this work.}
Madau et al.\ (1999) (and many other authors including Gnedin \& Ostriker 
1997) used the term ``reionization'' to label the overlap stage only, and
thus concluded that reionization was fast. Thus, all the semi-analytical 
models give more-or-less
similar predictions for the physics of reionization, and these
predictions are reproduced by numerical simulations.

\subsection{Reionization and the Lyman-alpha forest} 

While a question of reionization has its own astrophysical interest,
reionization also leaves an imprint on the IGM at lower
redshift, i.e.\ on the Lyman-alpha forest. One of the important
properties of the forest is the so-called ``effective equation of
state'', i.e.\ a tight relationship between the gas density and temperature
in the low density regime, which is usually can be approximated by a
power-law over a limited range of densities around the cosmic mean,
$$
	T \approx T_0(1+\delta)^{\gamma-1}
$$
(Hui \& Gnedin 1998). The time evolution of two parameters, $T_0$ and 
$\gamma$, depends on the thermal history of the universe, and therefore
is tightly coupled to the processes taking place during reionization.

Usually, the evolution of the equation of state is calculated using the
optically thin approximation and assuming that the radiation field is
uniform (Hui \& Gnedin 1998). This is clearly an approximation which needs
to be verified. 

\def\capES{
Evolution of the ``effective equation of state'' of the IGM, 
as represented by
parameters $T_0$ ({\it bottom panel\/}) and $\gamma$ ({\it top panel\/}),
for the full numerical simulation ({\it bold line\/}) and the
optically thin approximation ({\it thin line\/}). The shaded line marks
the redshift above which the temperature-density correlation is not well
established.
}
\placefig{
\begin{figure}
\insertfigure{\figdir/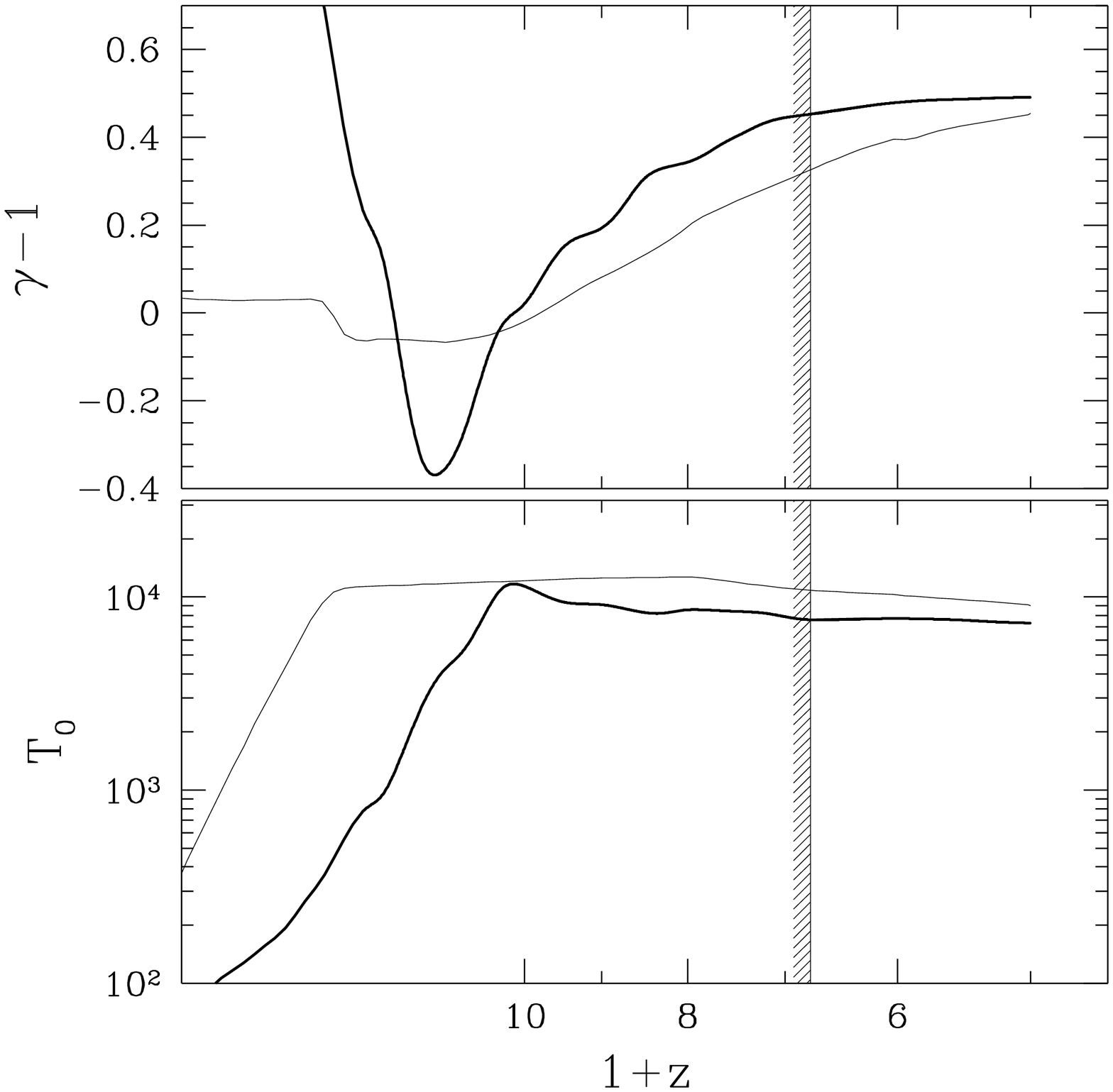}
\caption{\label{figES}\capES}
\end{figure}
}
Figure \ref{figES} shows the comparison of the ``effective equation of
state'' from the simulation and the one computed in the optically thin
approximation using the evolution of the spatially averaged
radiation field extracted from
the simulation (and thus both calculations have precisely the same
mean $J_\nu$ as a function of time). 
While the two calculations are quite different at high redshift,
the optically thin approximation gives a reasonably accurate answer at
$z\la 5$.

\def\capTD{
The joint mass-weighted 
distribution of the gas density and temperature
in the production run N128\_L4\_A at four different redshifts.
Bold solid lines show the power-law fits to the ``effective equation
of state'' at the respective redshifts.
}
\placefig{
\begin{figure}
\insertfigure{\figdir/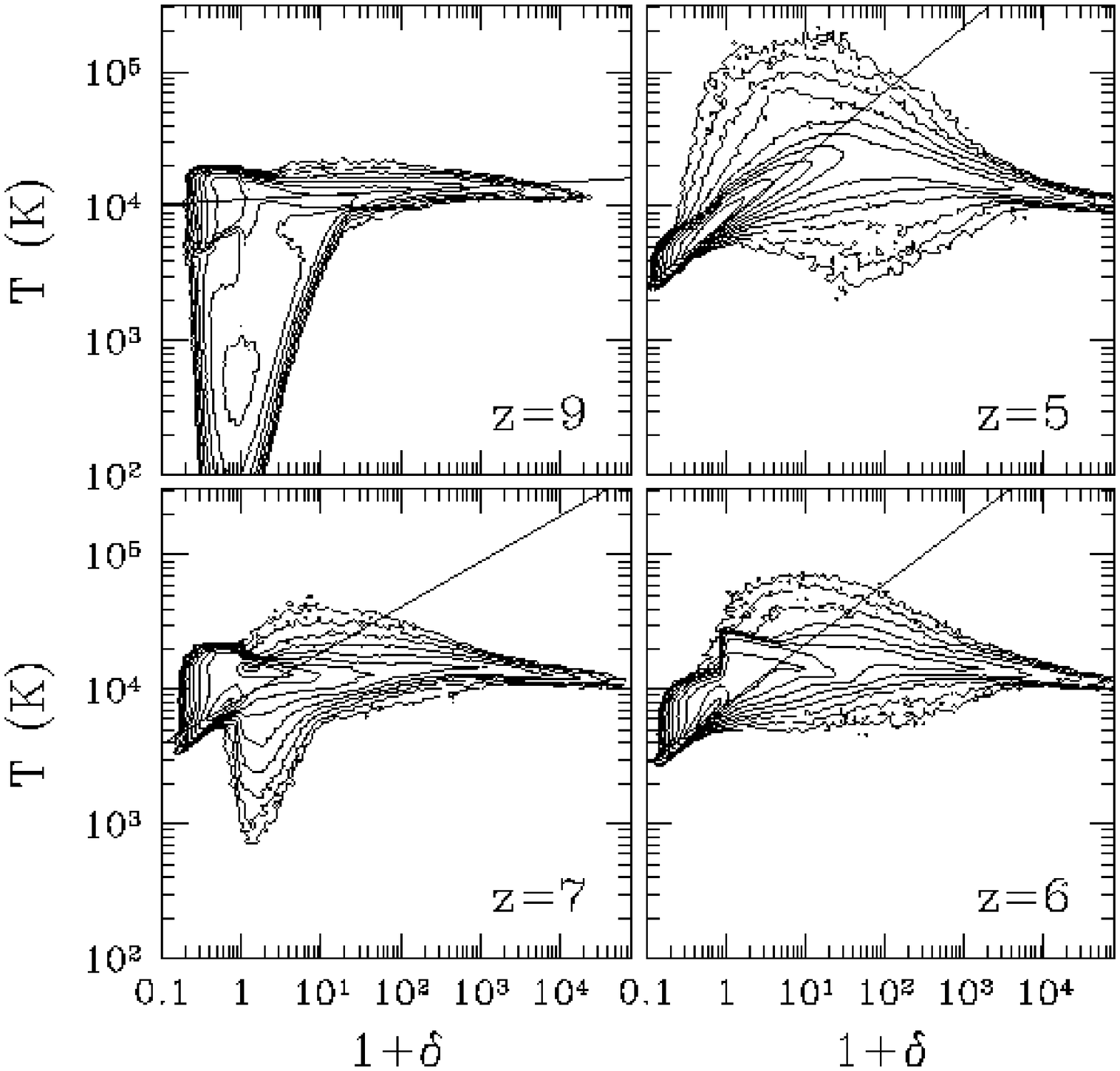}
\caption{\label{figTD}\capTD}
\end{figure}
}
It is however important to notice, that the relationship between the gas
density and temperature is more complicated at around the epoch of overlap and
before. Figure \ref{figTD} shows the joint mass-weighted distributions of
the gas density and temperature at four redshifts, corresponding to
Fig.\ \ref{figJD} and Fig.\ \ref{figXD}. One can see that the power-law
fits to the ``effective equation of state'' give poor representation of
the true temperature-density relation at $z\ga6$.

\def\capJR{Average ({\it thin lines\/}) and rms ({\it bold lines\/})
fluctuation in the ionizing background as a function of gas density at 
$z=5$ ({\it dotted lines\/}), 
$z=4.5$ ({\it dashed lines\/}), and
$z=4$ ({\it solid lines\/}).
}
\placefig{
\begin{figure}
\insertfigure{\figdir/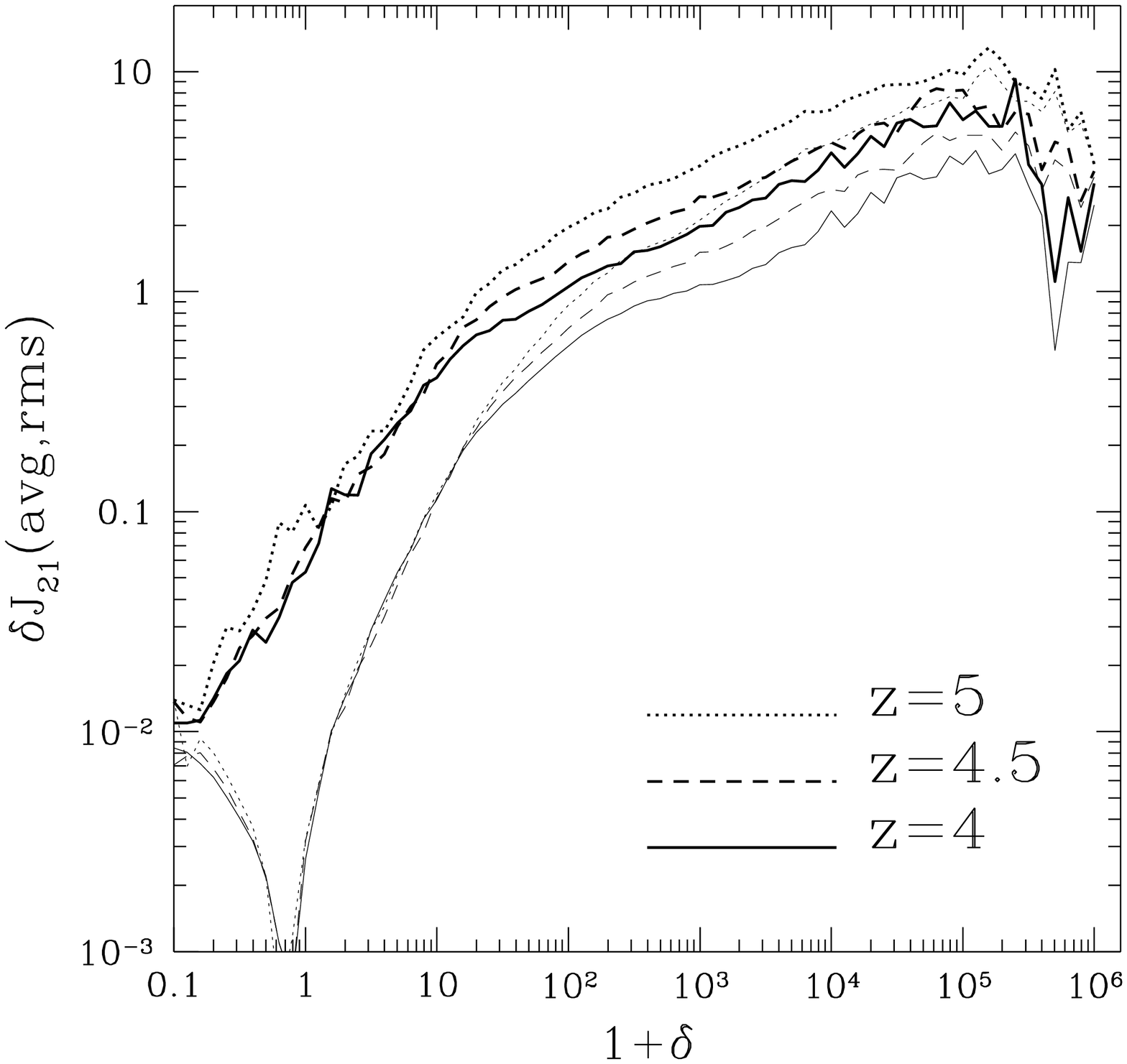}
\caption{\label{figJR}\capJR}
\end{figure}
}
Another important consideration that severely affects our ability
to model the Lyman-alpha forest is how homogeneous the ionizing background
is. Figure \ref{figJR} shows the mean and rms fluctuation in the local
ionizing background as a function of gas density at three values of redshift.
As can be seen, there is essentially no evolution in the spatial distribution
of the ionizing radiation in the low density regime, and gas with 
overdensities in excess of about 10 is subject to severe (more than 30\%)
inhomogeneities in the ionizing background. 

Since in the low density regime 
there exist a tight correlation between the column density of an
absorption line and the gas density the absorption line originates in,
it is possible to convert the gas density into the column density of the
absorption line. Assuming $J_{21}=0.5$ independently of redshift,
and using equation (18) of Ricotti, Gnedin, \& Shull (1999), I obtain:
$$
	N_\HI \approx 1.13\times10^{13}\dim{cm}^{-2}(1+\delta)^{1.52}
	\ \ \ \mbox{at}\ z=4,
$$
and
$$
	N_\HI \approx 2.57\times10^{12}\dim{cm}^{-2}(1+\delta)^{1.66}
	\ \ \ \mbox{at}\ z=3.
$$
Thus, if a 30\% fluctuation in the ionizing background is still compatible
with the background being uniform, only the Lyman-alpha forest with
$N_\HI<4\times10^{14}\dim{cm}^{-2}$ at $z=4$ and 
$N_\HI<1\times10^{14}\dim{cm}^{-2}$ at $z=3$
can be modeled in a uniform ionizing background approximation.
If an accuracy of 10\% 
is required (which corresponds to $\delta<1$), 
then these limits shrink further to 
$3\times10^{13}\dim{cm}^{-2}$ and $8\times10^{12}\dim{cm}^{-2}$ 
respectively.

It is important to remember, that the production run simulation
presented in this paper has still not converged numerically to
below several tens of percent accuracy, and thus the numbers quoted above
should be considered as only ``within-a-factor-of-two'' limits. Thus,
it is likely to be safe to claim that theoretical models that assume a 
homogeneous ionizing background cannot reproduce 
the Lyman-alpha forest 
with the
column density
$$
	N_\HI > 10^{15}\dim{cm}^{-2}\left(1+z\over 5\right)^6
$$
at a better than 30\% level, and the forest with 
$$
	N_\HI > 10^{14}\dim{cm}^{-2}\left(1+z\over 5\right)^6
$$
with a better than 10\% accuracy for $z$ between 3 and 4 on a
line-by-line basis.\footnote{Of course, these errors may average
out for some of the global properties of the forest.}

\section{Conclusions}

Cosmological numerical simulations that incorporate the effects of
radiative transfer show that the whole process of reionization
can be separated into three main stages: 
on a ``pre-overlap'' stage, which occupies a Hubble time or so, 
the $\HII$ regions expand into the neutral
low density 
IGM, leaving high density protrusions behind. The high density regions in the
vicinity of a source are being ionized as well, albeit with a much slower
rate. Pre-overlap culminates with the overlap, when the $\HII$ regions finally
merge and the universe becomes transparent on a much larger scale than
during the previous stage. The overlap epoch is characterized by a sharp
rise (in about 10\% of the Hubble time)
in the level of the ionizing background and in the mean free path. 
At the ``post-overlap'' stage the universe consists of highly ionized
low density gas, and still neutral high density gas, which is being
gradually ionized as more and more ionizing photons become available.
The boundary between the neutral and the ionized gas moves toward
higher densities with time.

In a currently fashionable CDM-type cosmological scenarios the whole process 
of reionization occupies a considerable part of the early evolution of
the universe from $z\sim15$ until $z\sim5$.

The aftermath of reionization has a profound effect on the IGM
at lower redshift: the residual fluctuations in the ionizing background
become quite significant at the column densities in the range from
$10^{14}\dim{cm}^{-2}$ to $10^{15}\dim{cm}^{-2}$ (at $z=3-4$) and may
compromise theoretical models that do not take these fluctuations into
account.

\acknowledgements

I am grateful to Jordi Miralda-Escud\'e and the referee Tom Abel
for valuable comments that significantly improved the original manuscript.
This work was partially supported by National Computational Science
Alliance under grant AST-960015N and utilized the SGI/CRAY Origin 2000 array
at the National Center for Supercomputing Applications (NCSA).

\appendix

\section{Local Optical Depth Approximation}

As was shown in Gnedin \& Ostriker (1997, equation [B9]),
the energy density of radiation
per unit frequency $I_\nu(x^i,t)$ can be represented as
\begin{equation}
	I_\nu(x^i,t) = \bar{I}_\nu(t) e^{\bar\tau_\nu(t)-\tau_\nu(x^i,t)}
	+ {a\over4\pi c}\int d^3x_1
	{S_\nu(x_1^i,t)-\bar S_\nu(t)\over(x^i-x_1^i)^2}
	e^{-\tau_\nu(x^i,x_1^i,t)},
	\label{inueq}
\end{equation}
where $\bar{I}_\nu(t)$ is the volume average of $I_\nu$, $S_\nu$ is
the source function, $\tau_\nu(x^i,x_1^i)$ is the optical depth between
$x^i$ and $x_1^i$, $\tau_\nu(x^i,t)$ is the optical depth from a given
point $x^i$ to an ``average'' point in the universe, obtained by solving
the homogeneous radiative transfer equation with no source function, and
$\bar\tau_\nu(t)$ is the normalization constant obtained from the following
condition:
$$
        e^{-\bar\tau_\nu(t)} \equiv \langle e^{-\tau_\nu(x^i,t)}\rangle
$$
with brackets denoting the volume average.

In order to implement a radiative transfer scheme into modern cosmological
simulations, one needs to be able to solve equation (\ref{inueq}) in
$O(N_B)$ operations (where $N_B$ is the number of baryonic resolution
elements, and I ignore a possible logarithmic multiplier, so that
$O(N_B\log N_B)$ still counts as O($N_B$)). However, equation (\ref{inueq}) 
in general 
includes two operations that are more expensive than this count:
(1) the one-point optical depth $\tau_\nu(x^i,t)$ requires $O(N_B\times N_D)$
operations, where $N_D$ is the number of directions, and in general is of
order of $N_B^{2/3}$, and (2) direct evaluation of the integral over the
source function requires $O(N_B\times N_S)$ operations, 
where $N_S$ is
the number of sources. In the case under consideration, $N_S$ is the number
of stellar particles in the simulations, and is typically
comparable to the number of baryonic resolution elements (perhaps, a factor
of a few smaller, but it is the scaling that matters). Thus, direct
(i.e.\ exact) implementation of equation (\ref{inueq}) is not feasible
at the moment. \footnote{In the case of reionization by quasars, however, 
when the number of sources
$N_S$ is small enough, the integral in (\ref{inueq}) can indeed be computed
directly in $O(N_B)$ operations (Abel et al 1999).}

Thus, one has to use approximations to try to decrease the operation count
in evaluating (\ref{inueq}). The ``Local Optical Depth'' approximation to 
evaluate the first term in (\ref{inueq}) was developed in Gnedin \& Ostriker
(1997) and is based on a simple observation that high precision in evaluating
$\tau_\nu(x^i,t)$ is not required: if the optical depth is small, it does
not matter much whether it is $10^{-3}$ or $10^{-5}$. If it is large, it 
again does not matter much whether it is $10^3$ or $10^5$. Only when
$\tau\sim 1$ it needs to be computed accurately, but the volume of space where
this condition is achieved is very small, so at the end no large error is
introduced even if $\tau$ is computed with only an order-of-magnitude
accuracy. Based on this consideration, the first {\it ansatz\/} is introduced
in the solution of equation (\ref{inueq}):
\begin{equation}
	\tau_\nu(x^i) = \sum_\alpha \sigma^{(\alpha)}_\nu n^{(\alpha)}(x^i) 
	L(x^i),
	\label{elldef}
\end{equation}
where index $\alpha$ runs over the list of species (in our case $\HI$,
$\GI$, and $\GII$), and $L$ is a characteristic length, which is taken to be
the characteristic length of the density distribution, and
I drop the time dependence (present in all terms) hereafter for the sake
of simplified notation.
Specifically, I adopt
the following expression for $L$:
\begin{equation}
        L = {1\over\sqrt{\alpha \left|\nabla\log\rho\right|^2 +
        \beta\rho\left|\Delta\log\rho\right| + 
	\gamma\left|\nabla\log x_{\HI}\right|^2}},
        \label{ellexp}
\end{equation}
where $\rho$ is the baryon density, $x_{\HI}$ is the neutral hydrogen
fraction, $\alpha=0.216$ and $\beta=0.068$ are constants 
\footnote{Note, that in Gnedin \& Ostriker (1997) these constants are
listed incorrectly because of a typo; they had proper values in the
simulations described in that paper.}
chosen to 
reproduce the correct result for the column density of a $1/(r^2+r_c^2)$
density distribution in two limits $r\rightarrow0$ and $r\rightarrow\infty$
and the last term is designed to work only for uniform density distribution,
when $L$ otherwise would be infinite. To this end I choose 
\begin{equation}
	\gamma = 30\max(0,1-\rho/\bar\rho),
	\label{gamdef}
\end{equation}
where $\bar\rho$ is the average density of the universe and the factor 30 in
front is chosen in order to satisfy the test D discussed above. Physically,
this coefficient means that the approximation described in this paper
spreads ionization fronts on average by a factor of $\sqrt{30}\sim5$.

The Local Optical Depth approximation ensures that the first term
in equation (\ref{inueq}) is computed in $O(N_B)$ operations.

Let me know turn to the second term. In general, integrals similar to the
one in (\ref{inueq}) can only be computed in $O(N_B)$ operations with
high quasi-Lagrangian resolution if they are convolutions. Thus, one of the
ways to make this integral computable is to approximate it as a sum of
several convolutions. For this purpose I divide all sources into two
categories, ``far'' and ``near'' sources, and rewrite equation
(\ref{inueq}) as
\begin{equation}
	I_\nu(x^i) = \bar{I}_\nu e^{\bar\tau_\nu-\tau_\nu(x^i)}
	+ I^F_\nu(x^i) 	+ I^N_\nu(x^i).
	\label{inueqfn}
\end{equation}
For a fluid element in a cosmological
simulation, a source can be considered to be ``far'' if it sits in a
separate clump. Then, the optical depth between the fluid element and the
source can be approximated as a sum of one-point optical depths from two
points:
$$
	\tau_\nu(x^i,x_1^i) \approx \tau_\nu(x^i) + \tau_\nu(x_1^i),
$$
and, thus,
\begin{equation}
	I^F_\nu(x^i) =  {a\over4\pi c}e^{-\tau_\nu(x^i)}\int d^3x_1
	{\left[S_\nu(x_1^i)-\bar S_\nu\right]
	e^{-\tau_\nu(x_1^i)}\over(x^i-x_1^i)^2}.
	\label{inueqf}
\end{equation}
This integral is a convolution, and can be computed in $O(N_B)$ operations
by means of a standard P$^3$M technique. More than that, since the SLH-P$^3$M
code already incorporates a P$^3$M gravity solver, it has been straightforward
to modify the existing solver to compute the (\ref{inueq}) term.

The second, ``near'' term, $I^N_\nu(x^i)$, can be considered to include
everything that is left to include in equation (\ref{inueqfn}) after
the approximation (\ref{inueqf}) is incorporated. However, in order to overcome
the $O(N_B\times N_S)$ count, I must introduce another ansatz.
For a ``near'' source, the two-point optical depth $\tau_\nu(x^i,x_1^i)$
can be decomposed into Taylor series,
$$
	\tau_\nu(x^i,x_1^i) \approx \nabla_j\tau_\nu(x^i,x^i)(x^j-x_1^j).
$$
However, even this is not enough since the factor
$\nabla_i\tau_\nu(x^i,x^i)$ still prevents the integral in (\ref{inueq})
from being a convolution. Thus, I introduce the second {\it ansatz\/}
in the following form:
\begin{equation}
	I^N_\nu(x^i) =  {a\over4\pi c}\int d^3x_1
	{S_\nu(x_1^i)-\bar S_\nu
	\over(x^i-x_1^i)^2}e^{-S|x^i-x_1^i|}\left[1-
	e^{-\tau_\nu(x_1^i)-\tau_\nu(x^i)}\right].
	\label{inueqn}
\end{equation}
Here $S$ is a constant, independent of position (but it still may depend on 
time), and the integral now is a sum of two convolutions.
The square bracket ensures that in the optically thin limit,
$\tau\rightarrow0$, the ``near'' term vanishes and the ``far'' term reduces to
$$
	I^F_\nu(x^i) =  {a\over4\pi c}\int d^3x_1
	{S_\nu(x_1^i)-\bar S_\nu\over(x^i-x_1^i)^2},
$$
which is indeed the integral in equation (\ref{inueq}) in the limit
$\tau\rightarrow0$. Thus, the proposed approximation {\it becomes exact}
in the optically thin limit.

The fact that the factor $S$ is the same for all sources implies that
the ionization front speed is correct only ``on average'', and each particular
ionization front may propagate with a speed different from the correct one,
but they will merge at approximately the right moment. Of course, if there
is only one ionization front in the simulation, its speed is computed
correctly.

From the argument presented above, it is clear that a good form for the factor
$S$ would be the following one:
\begin{equation}
	S(t) = C(t) \langle {\tau_{HI}(x^i,t)\over L(x^i,t)}\rangle,
	\label{sdef}
\end{equation}
where $\tau_{HI}(x^i)$ is the one-point optical depth at the hydrogen 
ionization threshold, and the ratio $\tau_{HI}/L$ represents the gradient
of the optical depth with the desirable feature that it is independent
of the poorly defined quantity $L$.

The quantity $C$, which in part determines the speed of ionization
fronts, needs to be fit to the tests. In other words, there is noting in
the approximation (\ref{inueqfn}-\ref{sdef}) which ensures photon number
conservation, so the quantity $C$ should be chosen so that this number
is at least approximately conserved.

Finally, a few words are in order about the frequency dependence.
For stellar sources, considered in this paper, the shape of the source
function is the same everywhere, so I can take it out of the integrals,
$$
	S_\nu(x^i,t) = g_\nu \rho_{MS}(x^i,t),
$$
where $g_\nu$ is constant in space and time, and $\rho_{MS}$ is the
mass density of massive stars. However, this still leaves the frequency
dependence of the optical depth in $I^F$ and $I^N$. With the current
computer capabilities it presents a considerable expense to perform
the calculation of the integral in $I^N$ and $I^F$ at a sufficient
number of frequency values.\footnote{Tests show that uniform logarithmic
sampling of the
frequency space should be at least 
50 points per decade, which makes the total number of required
frequency bins at least 200.}
Therefore, I introduce the third {\it ansatz\/} in the described
approximate scheme. I introduce three effective column densities
$N^{(\alpha)}_{\rm eff}$, where $\alpha$ again runs over $\HI$, $\GI$, and
$\GII$, so that
$$
	I^F_{\nu} = I^F_{\nu=0}\exp\left(-\sum_\alpha N^{(\alpha)}_{\rm eff}
	\sigma^{(\alpha)}_\nu\right),
$$
and analogously for $I^N$. Then the integrals in $I^F$ and $I^N$ only need
to be computed at zero frequency and at three threshold frequencies of
$\HI$, $\GI$, and $\GII$. In the future, when computer power increases by
another factor of 10 or so, it will be possible to get rid of the third
ansatz and compute integrals in $I^F$ and $I^N$ over a large number of
frequency bins.

\section{Testing the Approximation}

Obviously, the proposed approximation is so simple that it is highly unlikely
that it will work in all circumstances. Thus, my goal is to make sure that
it works in cosmologically relevant conditions. This can be achieved by 
appropriately ``training'' the scheme against spherically symmetric numerical
solutions, which can be obtained from direct integration of equation
(\ref{inueq}).

For the test, I choose the following density distribution:
\begin{equation}
	\rho(r) = {\rho_c r_c^n\over \left(r^3 + r_c^3\right)^{n/3}}.
	\label{rhotest}
\end{equation}
This density distribution peaks at the center with $\rho=\rho_c$, has
a core radius of $r_c$ (taken to be 11 comoving kiloparsecs in all cases),
and falls off as $r^{-n}$ outside of the core
radius. The specific form of the density law was chosen to make the
coordinate transformation from real space $x^i$ to Lagrangian space
$q^i$,
$$
	\rho d^3x = \bar\rho d^3q,
$$
analytical,
$$
	r = \left[\left({(3-n)\bar\rho q^3\over 3\rho_c r_c^n}+r_c^{3-n}
	\right)^{3/(3-n)}-
	r_c^3\right]^{1/3}.
$$
This distribution is then embedded into the uniform mesh $x^i=q^i$ at a point
$r_m$ where the average density inside the sphere with radius $r_m$ is 
$\bar\rho$,
so that $\langle \rho\rangle=\bar\rho$ over the computational box. 
Tests are then performed with the $64^3$ computational Lagrangian mesh
and compared to the results of a spherically symmetric calculation.
Spherically symmetric calculations are done with the same density profile
and with 4000 shells between $0.1\dim{kpc}$ to $100\dim{kpc}$. It has
been verified that this number of shells is sufficient to achieve a 
1\% accuracy in the exact solution.

\def\capIE{%
Distribution of the temperature ({\it top panel\/}) and neutral hydrogen 
fraction on linear ({\it middle panel\/}) and logarithmic 
({\it bottom panel\/}) scale for a homogeneous radiation field
test shining on the $n=2$ and $\rho_c/\bar\rho=10^3$ density
distribution from outside at $z=9$ 
with the photoionization rates of $\Gamma=10^{-12}\dim{s}^{-1}$
and $\Gamma=10^{-14}\dim{s}^{-1}$ taken at an arbitrary moments in time.
Solid lines show the exact spherically symmetric
solutions, and symbols mark the approximate 3D solutions, with solid 
squares corresponding to $\Gamma=10^{-12}\dim{s}^{-1}$ and open squares
corresponding to $\Gamma=10^{-14}\dim{s}^{-1}$.}
\placefig{
\begin{figure}
\epsscale{0.70}
\insertfigure{\figdir/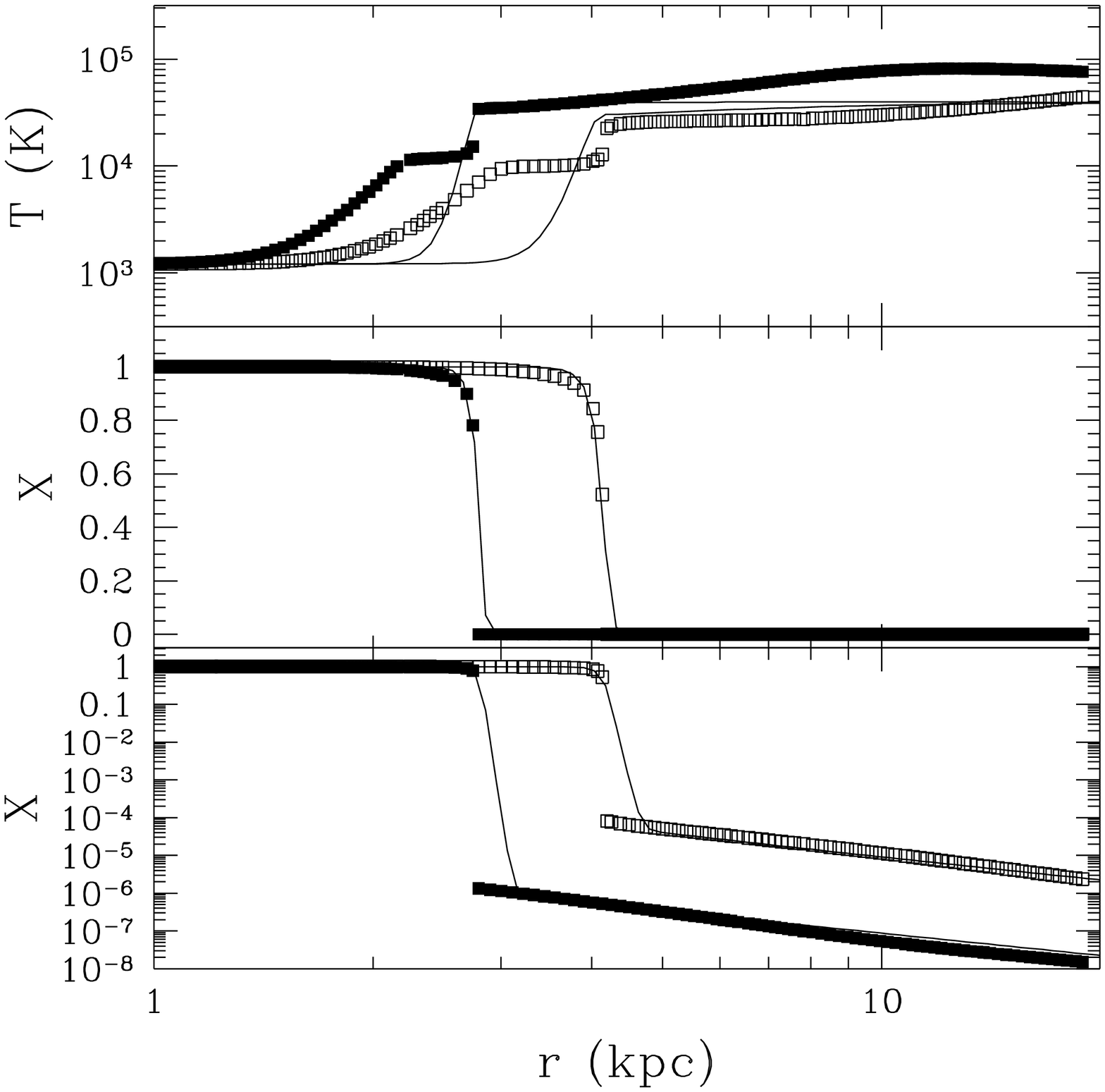}
\caption{\label{figIE}\capIE}
\end{figure}
}
As the first test, I adopt a $n=2$ ($\rho\propto r^{-2}$ ar large radii)
density distribution with $\rho_c/\bar\rho=10^3$ (maximum overdensity of
1000) embedded in the uniform radiation field with 
$\Gamma=10^{-12}\dim{s}^{-1}$ and $\Gamma=10^{-14}\dim{s}^{-1}$ (which roughly 
correspond to the radiation field after and before the overlap) at $z=9$.
No point
source is included in this test, so it designed to test only the expression
for the optical depth (\ref{elldef},\ref{ellexp}). Figure \ref{figIE} now shows
the temperature and the ionization fraction for the exact and the
approximate solutions taken at some arbitrary moments in time for the two
values of the photoionization rate. As one can see, the neutral hydrogen
profiles are reproduced reasonably well, except just after the ionization
front, whereas temperature profiles are reproduced less accurate, with
extra heating observed just after the ionization front. As I will show
below, the same behavior is observed in the test cases with the source
places at the center and the ionization front propagating outward. This
feature has to be considered an error of the LOD approximation and cannot
be easily repaired.

Next, I perform a series of test with a source of ionizing radiation
placed at the center of the spherically symmetric density
distribution. The strength of the source is parametrized by the photoionization
rate at the core radius $\Gamma_c$ computed in the optically thin
regime. 

\def\tabletests{
\begin{deluxetable}{cccccc}
\tablecaption{\label{tabtes} Parameters of Test Cases}
\tablehead{
\colhead{Case} & 
\colhead{$z$} & 
\colhead{$\Gamma_c a^3$ ($\dim{s}^{-1}$)} & 
\colhead{$\rho_c/\bar\rho$} & 
\colhead{$n$} & 
\colhead{$t_0$ (\dim{yr})} }
\startdata
A    &  4  & $10^{-14}$         & $10^3$ & $2$ & $1\times10^6$ \\
B    &  4  & $10^{-12}$         & $10^3$ & $2$ & $2\times10^4$ \\
C    &  4  & $10^{-10}$         & $10^3$ & $2$ & $4\times10^2$ \\
D    &  9  & $10^{-14}$         & $10^3$ & $2$ & $1\times10^6$ \\
E    &  9  & $10^{-12}$         & $10^3$ & $2$ & $2\times10^4$ \\
F    &  9  & $10^{-14}$         & $1$    & $-$ & $1\times10^6$ \\
G    &  9  & $10^{-12}$         & $40$   & $1$ & $1.5\times10^4$ \\
\enddata
\end{deluxetable}
}
\placefig{\tabletests}
The seven main tests performed are listed in Table \ref{tabtes}. 
The first five tests include the density distribution that falls
off as $r^{-2}$ at large radii, the test F verifies the propagation
of the ionization front over the uniform density field ($\rho_c/\bar\rho=1$),
and the last test G includes the $r^{-1}$ density law as the intermediate
case between the tests A-E and F. That test is not used to train the
approximate scheme, but rather to verify the performance of the approximation
in a substantially different arrangement.

Additional 
tests have been performed to verify numerical convergence and other technical
issues such as dependence on the cell size.

As the results of fitting the approximate scheme to the exact solutions of
the first six tests, the
following expression for the coefficient $C$ is derived:
\begin{equation}
	C = {1\over 0.1 + Q^{2/3}},
	\label{cdef}
\end{equation}
where
$$
	Q(t) = 8.6 \int_0^t {\tilde\Gamma(t^\prime)\over a(t^\prime)^{1/3}} 
	dt^\prime,
$$
and
$$
	\tilde\Gamma = \max\left(0,
	\langle \Gamma - {90\over C^{1/3}} R(T) n_e\rangle\right),
$$
where $\Gamma$ is the photoionization rate, $R$ is the recombination
coefficient, and $n_e$ is the electron number density. The quantity $C$ thus
depends on the time integral of the volume averaged photoionization rate
(with a correction for recombinations), and the factor $a^{1/3}$ in the
denominator accounts for the redshift of ionizing photons. 

Equation (\ref{cdef}) thus completes the approximation. I would like to stress
here that equation (\ref{cdef}) is a purely empirical fit and may have no
physical basis behind it. It appears physically plausible since it is based
on the number of ionizing photons emitted over the course of action, but
its specific form cannot be ``deduced'' from the first principles
(in large part due to the fact that the Local Optical Depth approximation
is based on two ansatzes).

\def\capIF{%
Evolution of the ionization front in seven test cases. The upper panel shows
cases A ({\it dotted line\/}), B ({\it short-dashed line\/}), and C 
({\it long-dashed line\/}). The lower panel shows cases
cases D ({\it dotted line\/}), E ({\it short-dashed line\/}), F 
({\it long-dashed line\/}), and G ({\it dot-dashed line\/}).
Thin lines mark the exact solution, and bold lines
show the Local Optical Depth approximation. Time axis is scaled arbitrarily,
with the scaling factor listed in Table \ref{tabtes}. The thin solid
line
shows the Shapiro analytical solution for the test F (homogeneous density
field).}
\placefig{
\begin{figure}
\epsscale{0.70}
\insertfigure{\figdir/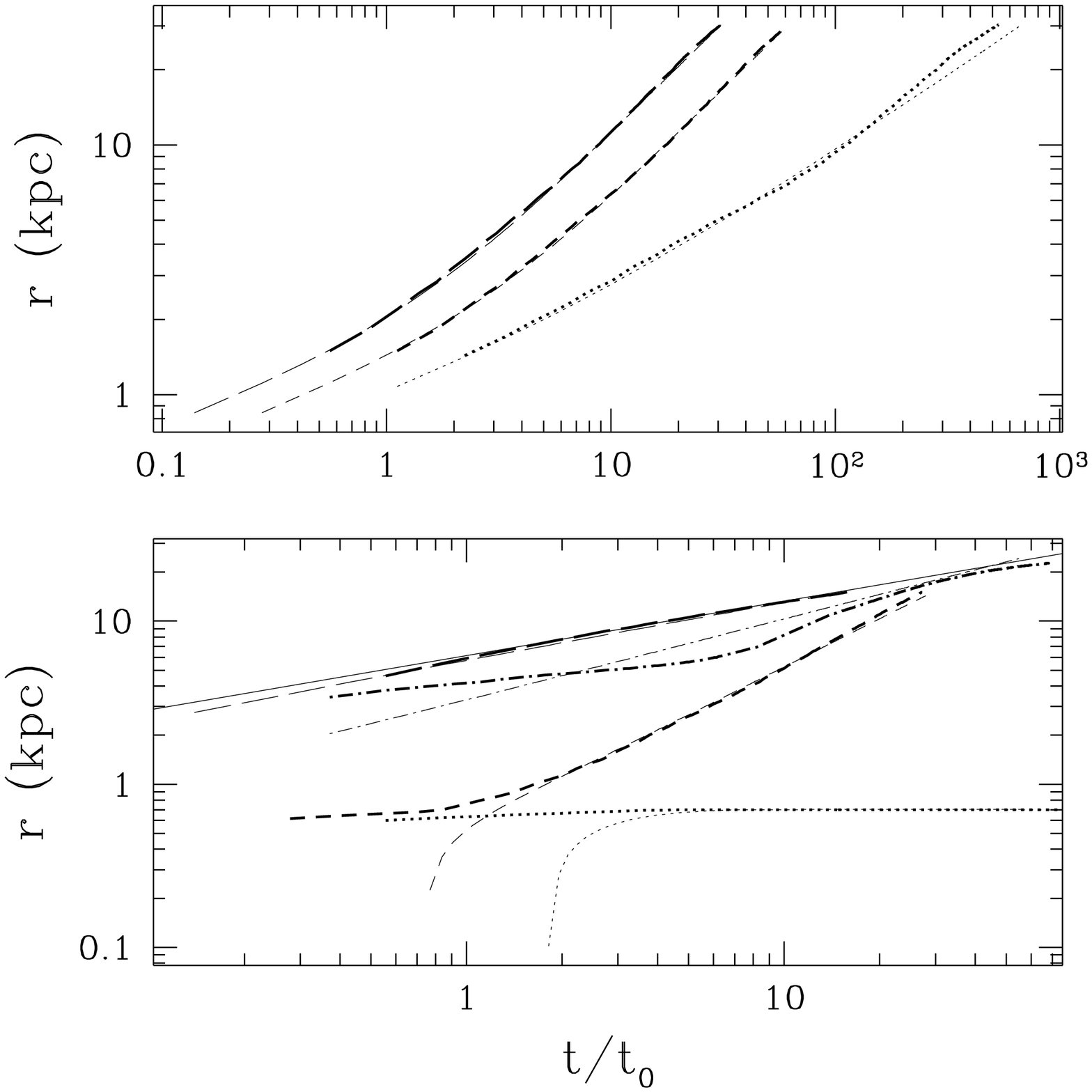}
\caption{\label{figIF}\capIF}
\end{figure}
}
We can now investigate how the developed approximation works in test cases.
Figure \ref{figIF} shows the comparison between the exact and approximate
calculations of the evolution of the ionization front in all seven test
cases. The agreement is good everywhere except cases D and E at early times,
but this is due to a finite resolution of a numerical simulation.
Deviations of the order of 20\% are also 
observed in case A at late times. The case G shows in general a worse 
agreement, because this test was not used to fit the approximate scheme
to the exact solutions. It can thus be used to demonstrate the level of
accuracy of the LOD approximation for arrangements different from those
used to ``train'' the approximation. The solid thin line in Fig.\ \ref{figIF}
shows the analytical approximation of Shapiro (1986) and
Shapiro \& Giroux (1987) for the case F (the homogeneous density field).
The small deviation observed at early times is due to the fact that the 
thickness of the ionization front in this regime is not negligible
compared to the size of the $\HII$ region, and the Shapiro solution becomes
invalid.

The question of numerical resolution in any 3D treatment of the radiative
transfer is of utmost importance. Gravitational collapse proceeds from
low density to high density, and therefore even with poor resolution
low density structures can be reproduced reliably. The situation is just
the opposite with the radiative transfer: the ionization front moves from
high density regions into low density regions. 
Thus, during the first time-step the
ionization front has to be fully resolved, or it will never leave the
resolution element it originated in (unless it is specifically followed
on inside that resolution element). Thus, to avoid
the problem of ionization front getting ``stuck'' due to lack of resolution,
I impose the condition that $S\epsilon>2$, where $\epsilon$ is the
gravitational softening length. In test cases E and F this leads to the
ionization front deviating from the exact solution at earlier times,
but catching up later, when it becomes fully resolved in the approximate 
scheme. 

\def\capIS{%
Distribution of the temperature ({\it top panel\/}) and neutral hydrogen 
fraction on linear ({\it middle panel\/}) and logarithmic 
({\it bottom panel\/}) scale
at five different times (as labeled by different lines)
in the exact solution ({\it thin lines\/}) and the approximate solution
({\it thick lines\/}) for cases 
A (panel {\it a\/}), 
D (panel {\it b\/}), and
B (panel {\it c\/}).}
\placefig{
\begin{figletters}
\begin{figure}
\inserttwofigures{\figdir/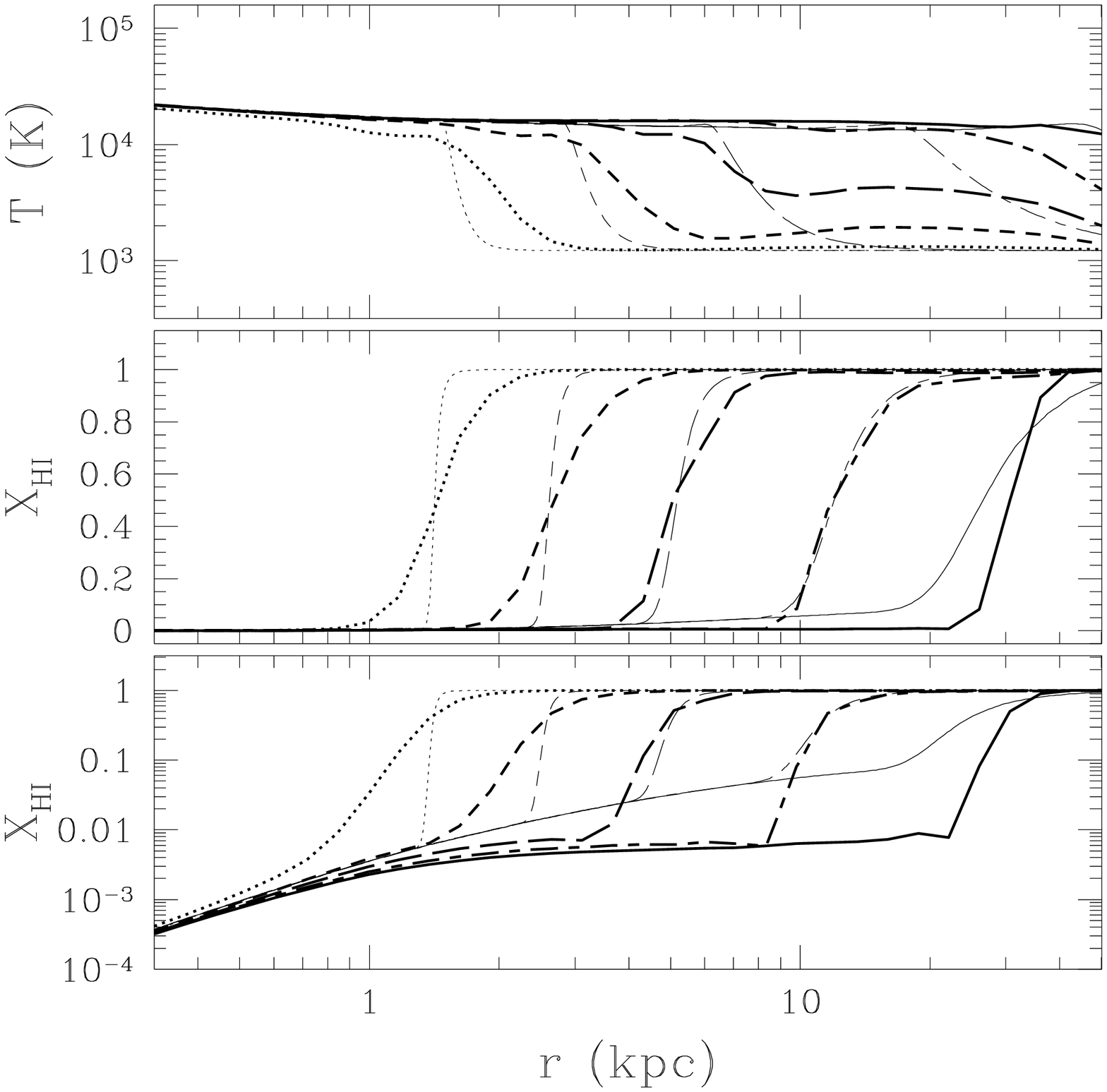}{\figdir/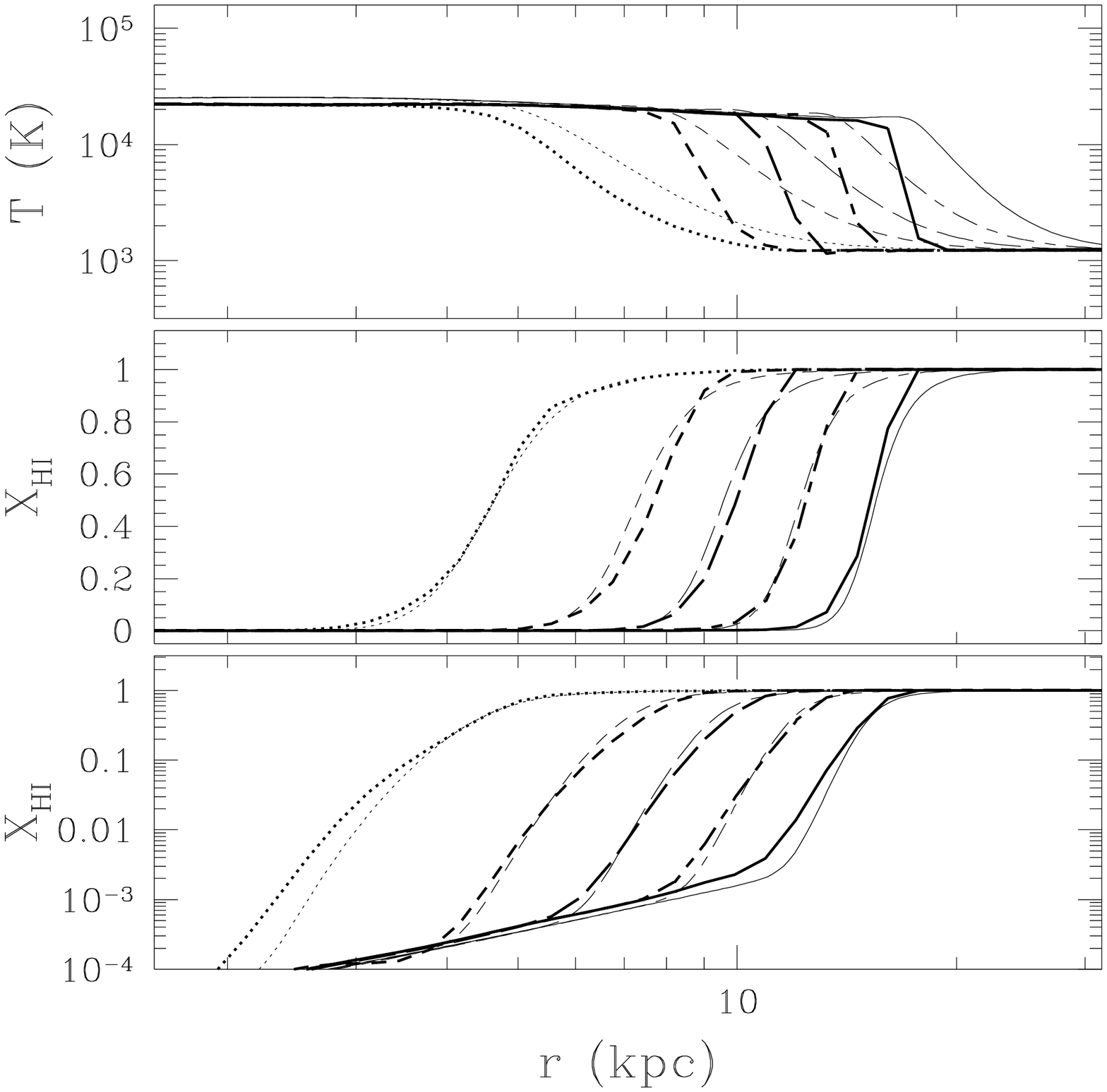}
\caption{\label{figIS}\capIS}
\end{figure}
}
Figure \ref{figIS} shows in three panels time-evolution of the temperature
and the neutral hydrogen fraction for cases A, B, and D. Since the
ionization front in the approximate solution is not sharp, I choose the
location where hydrogen is 50\% ionized as a location of the ionization
front in the approximate solution for the purpose of producing
Fig.\ \ref{figIF}. 

\def\capIC{%
Distribution of the temperature ({\it top panel\/}) and neutral hydrogen 
fraction on linear ({\it middle panel\/}) and logarithmic 
({\it bottom panel\/}) scale
in test B 
in the exact solution ({\it thin lines\/}) and the approximate solution
({\it thick lines\/}). Dotted lines show the full solution, and dashed
lines show the solution with cooling disabled. Cooling along the ionization
front leads to an incorrect prediction for the post-front temperature.}
\placefig{
\begin{figure}
\inserttwofiguresendfigletters{\figdir/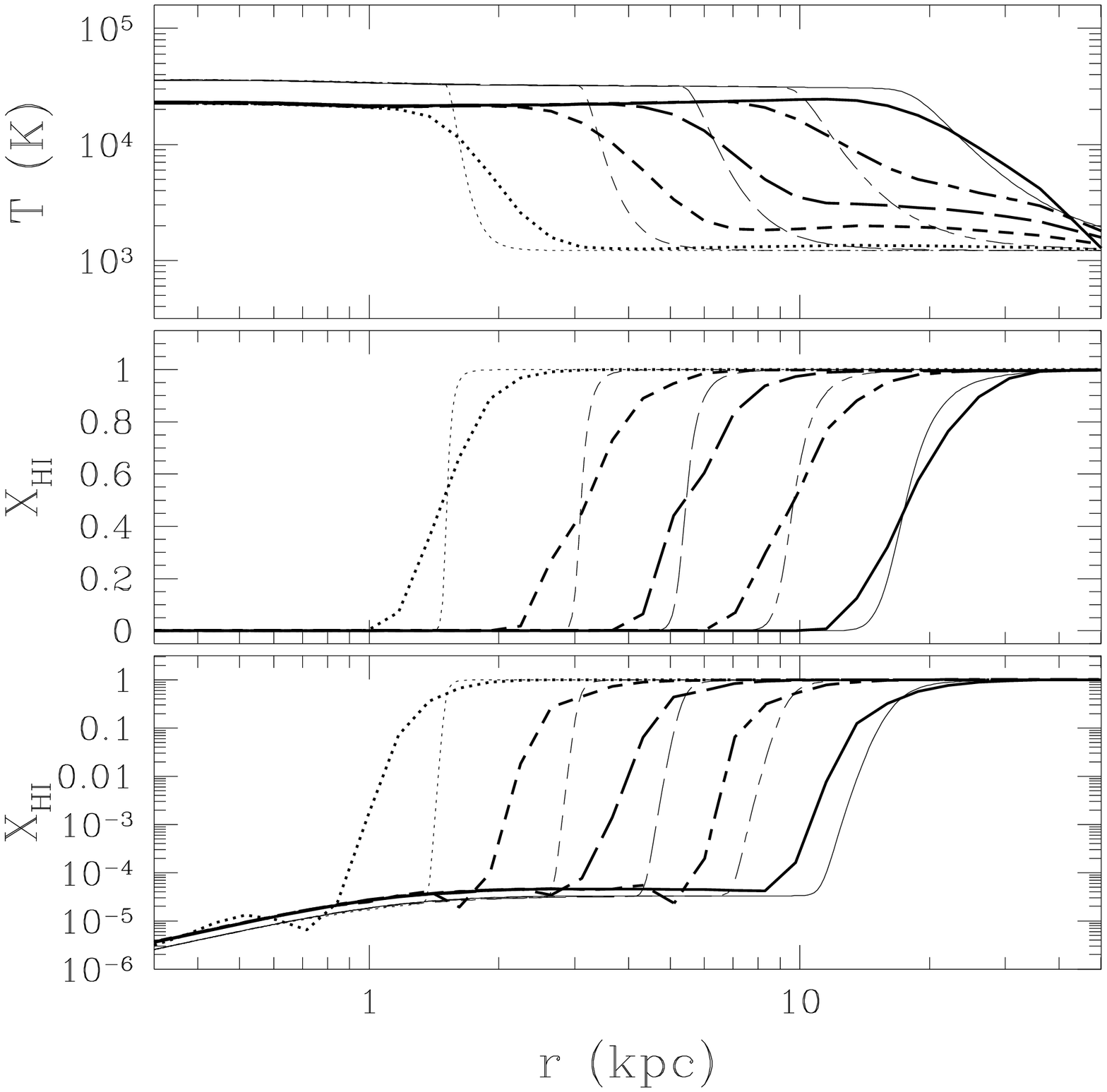}{\figdir/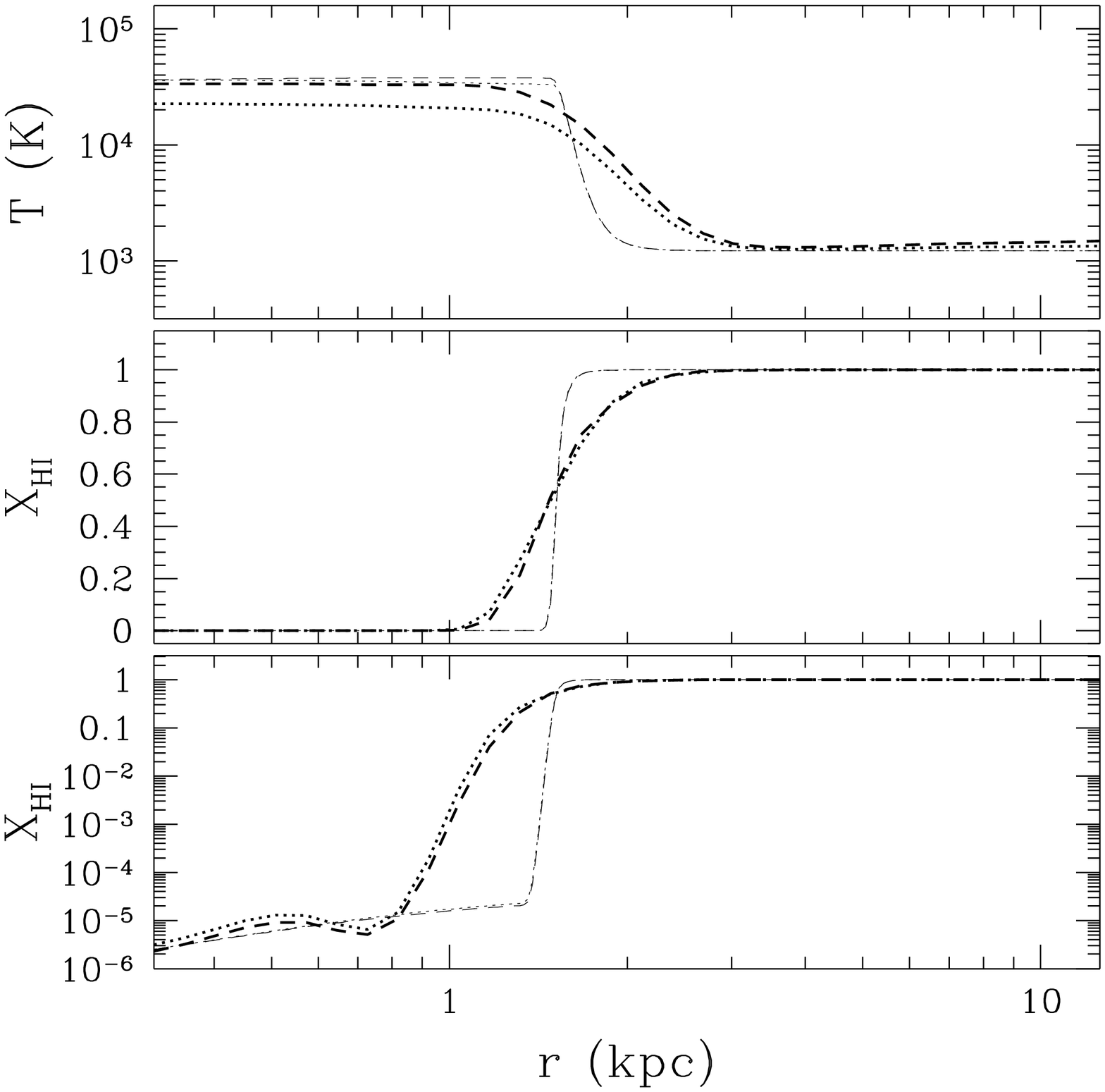}
\caption{\label{figIC}\capIC}
\end{figure}
\end{figletters}
}
Several defects of the approximate scheme are
apparent in Fig.\ \ref{figIS}. First, the ionization front is spread over
a considerable distance, but for cosmological purposes this distance is still
much smaller than the correlation length, and thus this does not make the big
impact on the dynamics of the gas. Second, at late times in Fig.\ \ref{figIS}a
the approximate scheme significantly overestimates the degree of ionization,
due to the fact that the optical depth between the position of the ionization 
front and the source is of the order of unity, and here the Local Optical
Depth approximation is bound to fail. Third, in Fig.\ \ref{figIS}a
the temperature ahead of the
front is higher than it should be, whereas in Fig.\ \ref{figIS}b it is
lower. This is due to the fact that leaking of high-energy photons across
the front is not captured correctly by the approximation. Finally,
the most severe defect is apparent in Fig.\ \ref{figIS}c: the temperature
inside the front in the approximate solution is some 50\% lower than it
should be. This is due to the fact that the ionization front is not sharp,
and the gas is able to cool while being ionized, whereas in the exact
solution the neutral fraction drops from 1 to $10^{-5}$ so quickly,
that no appreciable cooling occurs. 
To illustrate this further I show in Figure \ref{figIC} the first line from
Fig.\ \ref{figIS}b together with the exact and approximate solution for the
same test case but with disabled cooling. Without cooling, the approximate
solution predicts the post-front temperature right, but with cooling it allows
a considerable loss of energy.

This defect however is not specific to the Local Optical Depth approximation,
but obviously will be present in any scheme that does not resolve the 
ionization front. It is possible to modify the SLH technique in such a way as
to allow the mesh to deform appropriately and to achieve higher resolution at 
the ionization front, but this work is well beyond the scope of this paper.

Another significant defect of the Local Optical Depth approximation is the
absence of shadowing. Since LOD approximation does not include proper
ray tracing, the ionizing flux from a given source in a given
fluid element is only influenced by the opacity in the vicinity of the source
and the fluid element, and therefore a shadowing effect of a self-shielded
neutral cloud on the line of sight from the  fluid element to the source
would not be reproduced. Fortunately, shadowing is not important during
cosmological reionization: a self-shielded cloud with the size of 
$10\dim{kpc}$ moving with the velocity of $100\dim{km}/\dim{s}$ produces
a shadowing effect only during a time interval of $100\dim{Myr}$, which
is more than two orders of magnitude shorter than the recombination
time in the low density IGM. Thus, the ionization and the thermal state
of the IGM would not be affected by the passage of a cloud shadow.

\placefig{\end{document}}

\clearpage

\tableone

\clearpage

\tabletests

\clearpage

\newcounter{figurecap}
\setcounter{figurecap}{0}

\begin{center}
\bf Figure Captions
\end{center}

\refstepcounter{figurecap}
Fig.\ \thefigurecap---\label{figRS}\capRS

\refstepcounter{figurecap}
Fig.\ \thefigurecap---\label{figSF}\capSF

\refstepcounter{figurecap}
Fig.\ \thefigurecap---\label{figIM}\capIM

\refstepcounter{figurecap}
Fig.\ \thefigurecap---\label{figMP}\capMP

\refstepcounter{figurecap}
Fig.\ \thefigurecap---\label{figJD}\capJD

\refstepcounter{figurecap}
Fig.\ \thefigurecap---\label{figXD}\capXD

\refstepcounter{figurecap}
Fig.\ \thefigurecap---\label{figLA}\capLA

\refstepcounter{figurecap}
Fig.\ \thefigurecap---\label{figLF}\capLF

\refstepcounter{figurecap}
Fig.\ \thefigurecap---\label{figDE}\capDE

\refstepcounter{figurecap}
Fig.\ \thefigurecap---\label{figNG}\capNG

\refstepcounter{figurecap}
Fig.\ \thefigurecap---\label{figCF}\capCF

\refstepcounter{figurecap}
Fig.\ \thefigurecap---\label{figES}\capES

\refstepcounter{figurecap}
Fig.\ \thefigurecap---\label{figTD}\capTD

\refstepcounter{figurecap}
Fig.\ \thefigurecap---\label{figJR}\capJR

\refstepcounter{figurecap}
Fig.\ \thefigurecap---\label{figIE}\capIE

\refstepcounter{figurecap}
Fig.\ \thefigurecap---\label{figIF}\capIF

\refstepcounter{figurecap}
Fig.\ \thefigurecap---\label{figIS}\capIS

\refstepcounter{figurecap}
Fig.\ \thefigurecap---\label{figIC}\capIC


\begin{references}

\reference{ANM99}
Abel, T., Norman, M.\ L., \& Madau, P. 1999, \apj, 523, 66 

\reference{BG91}
Bertschinger, E., \& Gelb, J. 1991, Computers in Phys., 5, 154

\reference{BJ98}
Bond, J.\ R., \& Jaffe, A.\ H. 1998, Phil.\ Trans.\ Royal Soc.\ Lon.\ A,
in press (astro-ph 9809043)

\reference{CO99}
Chiu, W.\ A., \& Ostriker, J.\ P. 1999, in preparation

\reference{CF97}
Ciardi, B., Ferrara, A. 1997, \apj, 483, L5

\reference{CFGJ99}
Ciardi, B., Ferrara, A., Governato, F., \& Jenkins, A. 1999, \mnras,
submitted (astro-ph 9907189)

\reference{Gea94}
Giallongo, E., D'Odorico, S., Fontana, A., McMahon, H.\ G., Savaglio, S.,
Cristiani, S., Molaro, P., \& Treverse, D. 1994, \apj, 425, L1

\reference{GS96}
Giroux, M.\ L., \& Shapiro, P.\ R. 19996, \apjs, 102, 191

\reference{G95}
Gnedin, N.\ Y. 1995, \apjs, 97, 231

\reference{GB96}
Gnedin, N.\ Y., \& Bertschinger, E. 1996, \apj, 470, 115

\reference{GH98}
Gnedin, N.\ Y., \& Hui, L. 1998, \mnras, 296, 44

\reference{GO97}
Gnedin, N.\ Y., \& Ostriker, J.\ P. 1997, \apj, 486, 581

\reference{GBL98}
Griffiths, L.\ M., Barbosa, D., \& Liddle, A.\ D. 1998, \mnras, in press
(astro-ph 9812125)

\reference{HL97}
Haiman, Z., \& Loeb, A. 1997, \apj, 483, 21 

\reference{HL98}
Haiman, Z., \& Loeb, A. 1998, \apj, 503, 505

\reference{HG98}
Hui, L., \& Gnedin, N.\ Y. 1998, \mnras, 292, 27 

\reference {Lea96}
Lu, L., Sargent, W.\ L.\ W., Womble, D.\ S., \& Takada-Hidai, M. 1996, \apj, 
472, 509

\reference{M99}
Madau, P.\ 1999, invited review at the VLT Opening Symposium,
Antofagasta, Chile 1-4 March (astro-ph 9907268)

\reference{MHR}
Madau, P., Haardt, F., \& Rees, M. J. 1999, \apj, 514, 648

\reference{MMR}
Madau, P., Meiksin, A., \& Rees, M. J. 1997, \apj, 475, 429

\reference{MHR99}
Miralda-Escud\'e, J., Haehnelt, M., \& Rees, M. J. 1999, \apj, submitted 
(astro-ph/9812306)

\reference{NCO99}
Nagamine, K., Cen, R., \& Ostriker, J.\ P. 1999, \apj, submitted (astro-ph
9902372)

\reference{OG96}
Ostriker, J.\ P., \& Gnedin, N.\ Y. 1996, \apj, 472, L63

\reference{R99}
Renzini, A. 1999, in "When and How do Bulges
     Form and Evolve?", eds.\ C.M.\ Carollo, H.C.\ Ferguson, R.F.G.\ Wyse 
(Cambridge: Cambridge University Press), in press (astro-ph 9902361)

\reference{RGS99}
Ricotti, M., Gnedin, N.\ Y., \& Shull, J.\ M. 1999, \apj, submitted
(astro-ph 9906413)

\reference{S86}
Shapiro, P.\ R. 1986, \pasp, 98, 1014

\reference{SG87}
Shapiro, P.\ R., \& Giroux, M.\ L. 1987, \apj, 321, L107

\reference{SHCM99}
Songaila, A., Hu, E.\ M., Cowie, L.\ L., \& McMahon, R.\ G. 1999, \apj, 
525, L5

\reference{SAGDP99}
Steidel, C.\ C., Adelberger, K.\ L., Ciavalisco, M., Dickinson, M., \& 
Pettini, M. 1999, to appear in the proceedings of the Xth Rencontres de Blois,
"The Birth of Galaxies", July 1998 (astro-ph 9812167)

\reference{Tea96}
Tegmark, M., Silk, J., Rees, M.\ J., 
Blanchard, A., Abel, T., \& Palla, F. 1997,
\apj, 474, 1

\reference{VS99}
Valageas, P., \& Silk, J. 1999, \aa, in press (astro-ph 9907068)

\reference{Wea94}
Williger, G.\ M.\ et al. 1994, \apj, 453, L57

\end{references}
\end{document}